\documentclass{aa}
\usepackage[varg]{txfonts}
\usepackage{booktabs}
\usepackage{graphicx}
\usepackage{txfonts}
\usepackage{CJKutf8}
\usepackage{natbib}
\bibpunct{(}{)}{;}{a}{}{,}  
\usepackage{url}
\usepackage{longtable}
\usepackage{supertabular}
\usepackage{rotating}
\usepackage{multirow}
\usepackage{lscape}
\usepackage{color}
\usepackage{amsmath,amssymb,latexsym}
\usepackage{float}
\usepackage{adjustbox}
\usepackage{array}
\newcolumntype{+}{>{\global\let\currentrowstyle\relax}}
\newcolumntype{^}{>{\currentrowstyle}}

\usepackage[flushleft]{threeparttable}

  \font\aipsfont=cmsy10 at 10 pt                

\def\AIPS{{\aipsfont AIPS}}

\def\agn{AGN}

\def\cgm{CGM}

\def\hzrg{HzRG}
\def\hzrgs{HzRGs}

\def\sfr{SFR}

\def\imf{IMF}
\def\fwhm{FWHM}
\def\vla{VLA}
\def\pdr{PDR}

\def\galaxy{4C\,19.71}

\def\Lya{Ly$\alpha$}
\def\lya{Ly$\alpha$}

\newcommand{\CIV}{\mbox{C\,{\sc iv}}}
\newcommand{\CI}{\mbox{[C\,{\sc i}]}}
\newcommand{\CIII}{\mbox{C\,{\sc iii}]}}
\newcommand{\OIII}{\mbox{[O\,{\sc iii}]}}
\newcommand{\OII}{\mbox{[O\,{\sc ii}]}}
\newcommand{\OI}{\mbox{[O\,{\sc i}]}}
\newcommand{\HeII}{\mbox{He\,{\sc ii}}}
\newcommand{\HII}{\mbox{H\,{\sc ii}}}
\def\CIline{\mbox{[C\,{\sc i}]\,$^3$P$_1$-$^3$P$_0$}}
\def\CIlower{\mbox{[C\,{\sc i}](1--0)}}
\def\CIupper{\mbox{[C\,{\sc i}](2--1)}}
\def\CII{\mbox{[C\,{\sc ii}]}}
\def\CIIline{\mbox{[C\,{\sc ii}]\,$^2$P$_{3/2}$-$^2$P$_{1/2}$}}
\def\CI{\mbox{[C\,{\sc i}]}}
\def\Htwo{H$_2$}

\def\NII{\mbox{[N\,{\sc ii}]}}
\def\NIIline{\mbox{[N\,{\sc ii}]\,$^3$P$_1$-$^3$P$_0$}}

\def\CO13{\mbox{$^{13}$CO(4--3)}}

\def\kms{km\,s$^{-1}$}

\def\mstar{M$_\star$}

\def\msun{$M_{\odot}$}
\def\cmthree{cm$^{-3}$}

\def\mmol{M$_{\rm H_2}$}
\def\mion{M$_{\rm HII}$}

\def\mum{\,$\mu$m}
\def\msun{\,M$_{\odot}$}

\def\nodata{...}
\def\fgas{$f_{\rm gas}$}
\def\spitzer{{\it Spitzer}}
\def\herschel{{\it Herschel}}
\def\chandra{{\it Chandra}}

\def\mips1{MIPS (24\,\mum/)}

\usepackage[maxfloats=50]{morefloats}

\begin{document}

\title{ALMA and MUSE observations reveal a quiescent multi-phase circumgalactic medium around the z$\simeq$3.6 radio galaxy 4C~19.71\thanks{Based on observations obtained at the European Organization for Astronomical Research in the Southern Hemisphere under program 097.B-0323(B).}}

\author{Theresa Falkendal \inst{1,2,3}
  \and Matthew D. Lehnert\inst{1}
  \and Jo{\"e}l Vernet\inst{2}
  \and Carlos De Breuck\inst{2}
  \and Wuji Wang\inst{2}}

\institute{Sorbonne Universit\'{e}, CNRS UMR 7095, Institut d'Astrophysique de Paris, 98bis bvd Arago, 75014, Paris, France
  \and European Southern Observatory, Karl-Schwarzchild-Str. 2, 85748 Garching, Germany
  \and Potsdam Institute for Climate Impact Research (PIK), Member of the Leibniz Association, 14412 Potsdam, Germany}

\abstract{We present MUSE@VLT imaging spectroscopy of rest-frame ultraviolet emission lines and ALMA observations of the \CIline\ emission line, probing both the ionized and diffuse molecular medium around the radio galaxy \galaxy\ at z$\simeq$3.6. This radio galaxy has extended \Lya\ emission over a region $\sim$100\,kpc in size preferentially oriented along the axis of the radio jet.  Faint \Lya\ emission extends beyond the radio hot spots. We also find extended \CIV\ and \HeII\ emission over a region of $\sim$150\,kpc in size, where the most distant emission lies $\sim$40\,kpc beyond the north radio lobe and has narrow full width half maximum (\fwhm) line widths of $\sim$180\,\kms\ and a small relative velocity offset $\Delta$v$\sim$130\,\kms\ from the systemic redshift of the radio galaxy. The \CI\ is detected in the same region with \fwhm$\sim$100\,\kms\ and $\Delta$v$\sim$5\,\kms , while \CI\ is not detected in the regions south of the radio galaxy. We interpret the coincidence in the northern line emission as evidence of relatively quiescent multi-phase gas residing within the halo at a projected distance of $\sim$75\,kpc from the host galaxy. To test this hypothesis, we performed photoionization and photo-dissociated region (\pdr) modeling, using the code Cloudy, of  the three emission line regions: the radio galaxy proper and the northern and southern regions. We find that the \CI/\CIV\,$\lambda\lambda$1548,1551 and \CIV\,$\lambda\lambda$1548,1551/\HeII\ ratios of the two halo regions are consistent with a \pdr\ or ionization front in the circumgalactic medium likely energized by photons from the active galactic nuclei. This modeling is consistent with a relatively low metallicity, 0.03$<$[Z/Z$_{\odot}$]$<$0.1, and diffuse ionization with an ionization parameter (proportional to the ratio of the photon number density and gas density) of log U$\sim-$3 for the two circumgalactic line emission regions. Using rough mass estimates for the molecular and ionized gas, we find that the former may be tracing $\approx$2-4 orders of magnitude more mass. As our data are limited in signal-to-noise due to the faintness of the line, deeper [CI] observations are required to trace the full extent of this important component in the circumgalactic medium.
} 


\titlerunning{Combining MUSE and ALMA data}
\authorrunning{T. Falkendal et al.}

\date{Received 8 February 2019 / Accepted 8 July 2020}

\maketitle
 
\section{Introduction}
The circumgalactic medium (\cgm) is believed to control the gas supply of galaxies. It is the component of galaxies where recycled gas ejected from the galaxy by stars or active galactic nuclei (\agn s) likely mixes with accreted gas, whether from cosmological cold streams or gas accreted at the virial radius from the intergalactic medium. The \cgm\ is very diffuse, has low surface brightness emission lines, and is thus challenging to observe in anything other than absorption lines. Over the last three decades, many studies of the \cgm\ have focused on \agn s, observing their \cgm\ in \lya\ and other ultraviolet (UV) emission lines using narrow-band imaging or long-slit spectroscopy \citep[e.g.,][and sometimes H$\alpha$, e.g., \citealt{shopbell1999}]{Heckman1991a, Heckman1991b, Reuland2003, reuland2007}. Integral field spectrographs with relatively large fields of view, such as the multi-unit spectroscopic explorer (MUSE), PCWI, and KCWI, have recently made significant advances in our understanding of the spatial extent, ionization state, and dynamics of the \cgm\ in emission lines from ionized gas. Extended \Lya\ halos have recently been observed  with MUSE around a large fraction of star-forming galaxies at redshifts 3$\leq$z$\leq$6 in the \text{Hubble} Deep Field South and other regions \citep{Wisotzki2016, Leclercq2017, leclercq20, bielby20}. Extended \Lya\ halos have also shown kinematic signs, suggesting large-scale rotation of accreting material \citep{Prescott2015} and a filamentary structure \citep{Cantalupo2014,Vernet2017, martin19, umehata19}. \cite{ArrigoniBattaia2018} observed an enormous \Lya\ nebula around a radio-quiet quasar at z=3.164 with MUSE, spanning $\sim$300\,kpc and showing accretion of substructures onto the host quasar. With MUSE it is now possible to characterize the physical properties of the \cgm\ around galaxies. Observations have shown spectacular extended ionized regions around quasi-stellar objects (QSOs), quasars, radio galaxies, and star-forming galaxies. 

The \cgm\ around high-redshift radio galaxies (\hzrgs) can also contain large extended reservoirs of molecular gas (i.e.,$\sim$70-100\,kpc), which have been detected in \CIline, CO(1--0), CO(4--3), and H$_2$O \citep{Emonts2016, Gullberg2016, Emonts2018}, and sometimes in isolated regions of molecular emission \citep{Emonts2014,Gullberg2016_0943}. These isolated regions of emission contain dynamically quiescent gas with a relatively low velocity offset and could potentially be explained as being part of an accretion flow. \citet{Gullberg2016_0943} also note that the nuclear molecular gas in the radio galaxy MRC\,0943-242 is very narrow and dynamically quiescent compared to the typical broad optical and UV emission line gas seen close to the \agn s. In the case of one radio galaxy in particular, MRC\,1138-262, the amount of extended molecular gas estimated from these observations shows that it is sufficient to fuel the in situ star formation taking place in the \cgm. In fact, the properties of the star formation and molecular gas in the \cgm\ of the radio galaxy MRC\,1138-242 are such that the \cgm\ falls along the relationship between the star-formation rate (SFR) and molecular gas mass surface densities \citep[the ``Schmidt-Kennicutt relation'';][]{Emonts2016}. Molecular gas on even larger scales ($\sim$250\,kpc) has been interpreted as the denser regions of accretion streams feeding the central massive galaxies. Ongoing star formation in the streams may enrich the gas even before it is accreted, thus explaining the observations \citep{Ginolfi2017}. In addition, circumgalactic molecular gas around some \agn s\ has been interpreted as being due to outflows \citep[e.g.,][]{Cicone2015}. These observations support the view that the \cgm\ is in part metal-rich and dense, providing a link between cold halo gas, any potential in situ star formation, and supplying gas to the host galaxy \citep[see e.g.,][]{Emonts2016, Vernet2017, ArrigoniBattaia2018, martin19}. 

The low-J CO emission lines have proven to be good tracers of diffuse, low-density molecular gas. Several studies suggest that \CI\ can be an equally good tracer, complementing the other tracers of the total molecular content of gas \citep{Papadopoulos2004, Glover2015}. The \CI\ lines have similar critical densities to those of the lower rotational transitions of CO and likely trace the same diffuse gas. But unlike the optically thick low order transitions of CO, \CI\ is optically thin and therefore probes higher column densities than CO \citep{Papadopoulos2004}. Moreover, its low critical density implies that \CI\ does not probe the densest gas, which may ultimately limit its utility as a tracer of high column densities of H$_2$. Recent studies have also investigated if CO in star-forming galaxies might effectively be destroyed by cosmic rays while leaving \Htwo\ intact \citep{Bisbas2015,Bisbas2017}. The results of these studies suggest that \CI\ might be a more reliable \Htwo\ tracer in environments of intensely star-forming galaxies, \agn s, and near regions of synchrotron emission such a radio jets and lobes, where the cosmic ray intensity may be high relative to the Milky Way and other nearby normal galaxies \citep{Papadopoulos2018}.

Galaxy \galaxy\ is a massive, double-lobed, steep-spectrum, FR-II type \hzrg\ with X-ray emission seen at both the core and hot spots over an extent of $\sim$60\,kpc \citep{Smail2012}. The galaxy has extended \OIII$\lambda$5007 line emission spanning from the host out to both the lobes \citep{Armus1998}. The large emission size of 66\,kpc$\times$16\,kpc is unusual compared to radio galaxies at $z\sim$2 \citep{Nesvadba2017}. The galaxy has bright extended \Lya, \CIV,\ and \HeII\ halos of the size of $\sim$120\,kpc, observed with long-slit spectroscopy \citep{Maxfield2002}. Galaxy \galaxy\ is a unique galaxy; it is bright across a wide wavelength range and the face-on orientation together with a luminous \agn\ provide a test-bed with which to study the cold molecular and warm ionized gas within the galaxy, the interaction of the radio jets with the ambient medium as they expand outward, and the properties of halo gas.

Throughout the paper we assume a flat $\Lambda$ cold dark matter ($\Lambda$CDM) cosmology with $H_0=67.8$\, \kms\,Mpc$^{-1}$, $\Omega_{\rm M}$=0.308, and $\Omega_{\Lambda}$=0.692 \citep{PlanckCollaboration2016}, which implies a scale of 7.427\,kpc\,arcsec$^{-1}$ at $z=3.59$.

\section{Data}

\subsection{ALMA observations}

The Atacama Large Millimeter/submillimeter Array (ALMA) cycle 3 observations in band 3 were carried out on UT 2016 March 6, with 38 min on-source integration and 38 operating antennas. The four 3.875\,GHz wide bands were tuned to include the \CIline\ (\CIlower\ hereafter) and $^{13}{\rm CO}\,J=4 \rightarrow 3$ lines, covering 94.6--98.4\,GHz and 106.5--110.3\,GHz, respectively. We calibrated the data with the Common Astronomy Software Applications (CASA) using the observatory-supplied calibration script. The data cube and moment-0 maps were also produced using CASA. A natural weighting (robust parameter of 2) was applied since the data has low signal-to-noise. This results in a continuum image with synthesized beam 1.8"$\times$1.9" and a root mean square (RMS) of 14\,$\mu$Jy. To produce the cube and look for \CIlower\ ($\nu_{\rm rest}$=492.16\,GHz) and \CO13 ($\nu_{\rm rest}$=440.77\,GHz) emission lines, we binned the data to 50\,km~s$^{-1}$, which resulted in an RMS noise of 0.3\,mJy. To create the moment-0 maps, we first subtracted the continuum in the $uv$-plane by fitting a first-order polynomial over all the spectral window (but excluding the channels where the \CIlower\ and \CO13\ emission were expected to lie). We then collapsed the cube over the frequency range  107.21--107.27\,GHz and 96.007--96.067\,GHz for the \CIlower\ and \CO13\ moment--0 maps, respectively, resulting in line-only continuum-free images.

\subsection{MUSE observations}

The MUSE observations with the Very Large Telescope (VLT), Unit Telescopes 4 (program ID 097.B-0323(B)) were carried out over four nights between 2016 June 7 and 2016 September 2, with a total of five hours on source integration time during which the average seeing was $\sim$1.2". The data were taken in five sets of two 30\,min exposures rotated 90 degrees with respect to one another, resulting in a total of ten exposures. We reduced the data using the MUSE instrument pipeline version 1.6.2. First, we preprocessed the data with the standard settings to produce a calibrated and sky-subtracted combined data cube. But due to strong sky lines that were not optimally subtracted  and thus created artifacts in the cube (especially the \OI$\lambda$5577 night sky line which lies coincident with the redshifted wavelength of \Lya\ emission from \galaxy), we decided to reprocess the cube. To overcome the problems with the previous reduction, we reduced each exposure individually without sky subtraction and put them on the same spatial and spectral coordinate grid by providing the first exposure (astrometry corrected) as the \textsc{OUTPUT\_WCS} argument for the other nine exposures. Then, we sky-subtracted each exposure by using ZAP 2.0 \citep[the Zurich Atmosphere Purge;][]{Soto2016}. We ran ZAP with the default parameters and since it was run on individual exposures, the contribution from the faint sources of emission in the data had a negligible impact on the sky subtraction. Finally, we merged all ten individual 30\,min exposures into one combined cube using the python package MPDAF 2.5 (MUSE python data analysis framework) developed by the MUSE Consortium\footnote{https://mpdaf.readthedocs.io/en/stable/credits.html}. 

Moment-0 maps were created by subtracting the continuum emission by, in turn, fitting a first-order polynomial to the red and blue side of each emission line and then collapsing the cube over $\lambda_{obs}$=5581.64--5584.14\,\AA, $\lambda_{obs}$=7100.39--7102.89\,\AA, $\lambda_{obs}$=7525.39--7535.39\,\AA,\ and $\lambda_{obs}$=8739.14--8756.64\,\AA\ for  \Lya, \CIV,  \HeII, and  \CIII, respectively.

\subsection{Previous supporting observations}

\begin{figure*}[ht]
\centering
\includegraphics[width=\linewidth]{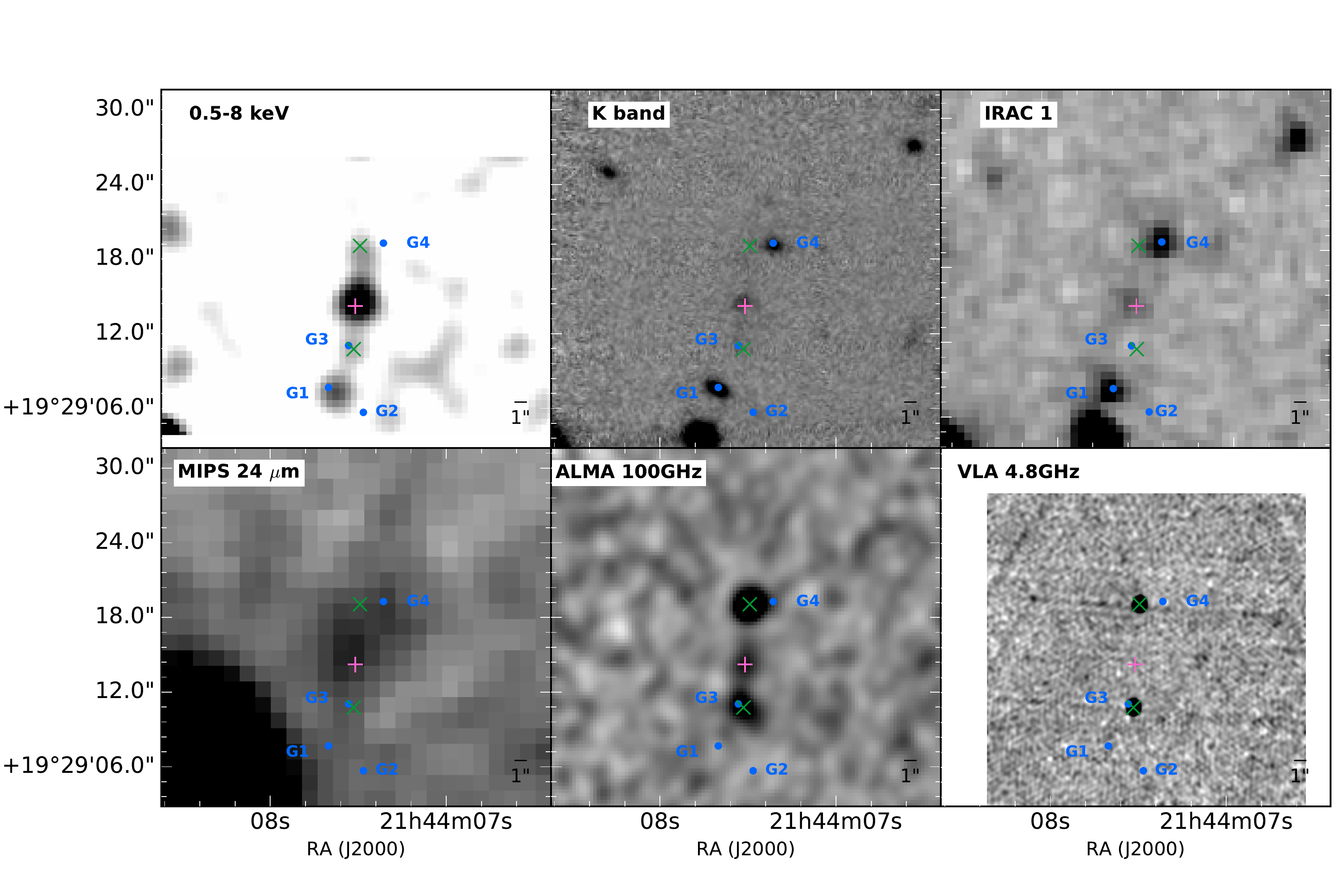}
\caption{Six views around \galaxy\ using previously published data. \textit{Top left:} smoothed 0.5--8\,keV \textit{Chandra} images \citep{Smail2012}; \textit{top center:} K-band 2.0--2.45\,\mum\ \citep{Armus1998}; \textit{top right:} IRAC 3.6\,\mum\ \citep{Seymour2007}; \textit{bottom left:} MIPS 24\,\mum\ \citep{DeBreuck2010}; \textit{bottom center:} ALMA band 3 continuum of 94.6--98.4\,GHz and 106.5--110.3\,GHz \citep{Falkendal2019}; \textit{bottom right:} \vla\ band C 4.8\,GHz \citep{Pentericci2000}. The pink plus signs indicate the center of the host galaxy determined from the peak of the thermal dust emission in the ALMA band 3 continuum image. Green crosses indicate the positions of the hot spots of the two radio lobes. The filled blue circles show the location of four foreground galaxies around \galaxy. For details about coordinates and redshifts of the sources in the field, see Table~\ref{tab:coordinates}.} 
\label{fig:support_data}
\end{figure*}

Galaxy \galaxy\ has an extensive set of observations, spanning from X-ray to radio frequencies (Fig.~\ref{fig:support_data}). The galaxy has been observed at radio wavelengths with the Very Large Array (\vla) bands C, X, and L \citep{Pentericci2000, Reuland2003}. Bands C and X show a clear and quite symmetric double lobe structure, but the core is not detected in any of the \vla\ bands. \chandra\ observations of \galaxy\ show that it has weak X-ray emission over $\sim60$\,kpc scales extending from the host galaxy and in the direction of the radio jet; the observations also show emission directly from the radio lobes \citep{Smail2012}. Galaxy \galaxy\ was observed through a K-band and narrow-band filter centered on the \OIII\ 5007\,\AA\ emission, which shows extended \OIII\ emission over a region of 74$\times$9\,kpc extended along the axis of the radio jets \citep{Armus1998}. Galaxy \galaxy\ was also part of a large \spitzer\ and \herschel\ survey of 70 \hzrgs\ $1<z<5.2$ \citep{Seymour2007, DeBreuck2010, Drouart2014} observed with the Infrared Array Camera (IRAC), the Infrared Spectrograph (IRS), and the Multiband Imaging Photometer (MIPS) on \spitzer, as well as the Photodetector Array Camera and Spectrometer (PACS) and the Spectral and Photometric Image Receiver (SPIRE) on \herschel.  The IRAC 3.6\mum, 4.5\mum,\ and MIPS 24\mum\ show continuum emission at the location of the host galaxy. Through spectral energy distribution (SED), fitting the stellar mass has been estimated to be log\,(\mstar/\msun)=11.13 \citep{DeBreuck2010} and an SFR of 84\,\msun\,yr$^{-1}$ \citep{Drouart2014, Falkendal2019}. The K-band image shown in Fig.~\ref{fig:support_data} is publicly available\footnote{http://www.eso.org/$\sim$cdebreuc/shzrg/} but unfortunately did not have an astrometric solution. We used \AIPS\ (Astronomical Image Processing System) and the task {\tt XTRAN} to determine the coordinate transformation between image pixels and the RA and Dec of reference stars. We used Gaia DR2 \citep{GaiaCollaboration2018} to retrieve the coordinates of the five brightest stars in the field, and, via {\tt XTRAN}, we obtained the astrometric solution for the K-band image.

\section{Results}

Combining ALMA submillimeter (submm) observations with MUSE optical observations reveals extended line emission at larger scales around \galaxy\ than previously observed. Figure~\ref{fig:support_data} shows X-ray 0.5--8\,keV continuum, K-band, IRAC 3.6\mum, MIPS 24\mum, ALMA 100\,GHz continuum, and \vla\ 4.8\,GHz images of \galaxy. The \agn\ and its host galaxy are detected in all of these wavelength bands except with the \vla, where only the two synchrotron lobes extending toward the north and south were detected. The \Lya\ image constructed from the MUSE data cube shows bright emission at the position of the host galaxy extending out toward both the northern and southern radio hot spots (Fig.~\ref{fig:Narrow_band}). Weak \Lya\ emission is also detected south of the southern lobe (region C, Fig.~\ref{fig:CIV_with_extracted_regions}). Similar morphological structure is also seen in \CIV, \HeII,\ and \CIII\ (Fig.~\ref{fig:Narrow_band}). The general structure and extent of the ionized gas is in agreement with previous long-slit spectroscopy observations \citep{Maxfield2002}. \CIlower\ line emission is detected at the position of the host galaxy (region B, Fig.~\ref{fig:CIV_with_extracted_regions}) and 9\,arcsec ($\sim$75\,kpc in projection) north of the source (region A, Fig.~\ref{fig:CIV_with_extracted_regions}). This extended \CIlower\ emission coincides with weak \CIV\ detected with MUSE (Fig.~\ref{fig:Narrow_band}). Region A is not detected in the submm continuum emission with a 3$\sigma$-upper limit of 0.4\,$\mu$Jy. Weak \CIV\ is also detected south of the source (region B), showing  velocity dispersion and velocity offset similar to the emission from region A. Figure~\ref{fig:CIV_with_extracted_regions} shows the three different regions from which the emission line spectra are extracted: A, north of the northern radio lobe; B, the host galaxy and \agn; and C, south of the southern radio lobe. In the following sections, we will discuss each of these regions in detail. 

\subsection{Molecular and atomic emission lines}
\label{sec:alma_lines}

We searched for \CIlower\ and \CO13\ lines in the ALMA cube by binning to different velocity widths and stepping through the velocity channels to search for emission (a ``blind search''). Due to the low signal-to-noise, we did not bin the cube for resolutions coarser than 50\,\kms. We found two \CIlower\ detections, one over region A and one in region B. The detection in the latter coincides with the peak position of the K-band (IRAC 3.6\,\mum\ and 4.5\,\mum), IRS 16\,\mum, and the ALMA thermal dust emission of the host galaxy and \agn\ (Fig.~\ref{fig:support_data}). From the moment-0 map of the \CIlower\ emission (summed over $\nu_{\rm obs}$=107.21--107.27\,GHz of the $uv$ continuum subtracted cube; Fig.~\ref{fig:Narrow_band}), we defined the boundaries of the two \CIlower\ regions (Fig.~\ref{fig:CIV_with_extracted_regions}). We extracted the spectra of region B from the $uv$ continuum subtracted ALMA cube and spectra of regions A and C from the non-$uv$ continuum subtracted ALMA cube. We extracted the data from region C, even though it is clearly not detected in \CIlower,\ simply to provide an estimate of the level of noise in these data. The \CO13\ line is not detected with ALMA. A moment-0 map of the \CO13\ was constructed with the same width as the \CIlower\ moment-0 map ($\nu_{\rm obs}$=96.007--96.067\,GHz) and centered at $\nu_{\rm obs}$=96.037\,GHz, assuming the systemic redshift. The map reveals no detections and again indicates the noise level in these data.

We fit both \CIlower\ lines with a Gaussian profile to estimate the integrated line flux, line width, and velocity relative to systemic. We used the non-linear least square method to fit a Gaussian function to the spectrum using the RMS=0.3\,mJy as one sigma uncertainty in the flux. The reported errors are the square root of the variance, one sigma, of the parameter estimate for the Gaussian fits. We used the \CIlower\ line at the core, region B, to determine the systemic redshift, $z_{\rm sys}$=3.5895, of the host galaxy. The fitting results in an \fwhm\ of 87$\pm$23\,\kms\ and an integrated flux of (0.40$\pm$0.15)$\times$\,$10^{-18}$\,erg\,cm$^{-2}$\,s$^{-1}$. The extended detection in region A is only shifted $\sim$5\,\kms\ from the \CIlower\ at the core; it has an \fwhm\ of 108$\pm$54\,\kms\ and an integrated flux of (0.19$\pm$0.12)$\times$\,$10^{-18}$\,erg\,cm$^{-2}$\,s$^{-1}$. Figure~\ref{fig:MUSE_ALMA_lines} shows the spectra of the two detected \CIlower\ lines with the best fitting model, as well as a noise spectrum extracted from region C. Figure~\ref{fig:MUSE_ALMA_lines} also shows the noise spectrum of the \CO13\ line at the location of the host galaxy (region B). Table~\ref{table:Line_fluxes} lists the fitted parameters.

The optically thin $^{13}$CO lines are intrinsically faint (line ratio $^{12}$CO/$^{13}$CO$\sim$20--40 and \CIlower/$^{12}$CO(4--3)$\sim$0.6 \citep{Alaghband-Zadeh2013, Bothwell2017}) and have only been detected in a few high-$z$ galaxies \citep{Henkel2010,Danielson2013,Spilker2014,Zhang2018}. Our lack of a detection of $^{13}$CO(4--3) was expected and will not be discussed further. 

\subsection{Ionized gas}
\label{sec:muse_lines}

\begin{figure*}
\centering
\includegraphics[width=\linewidth]{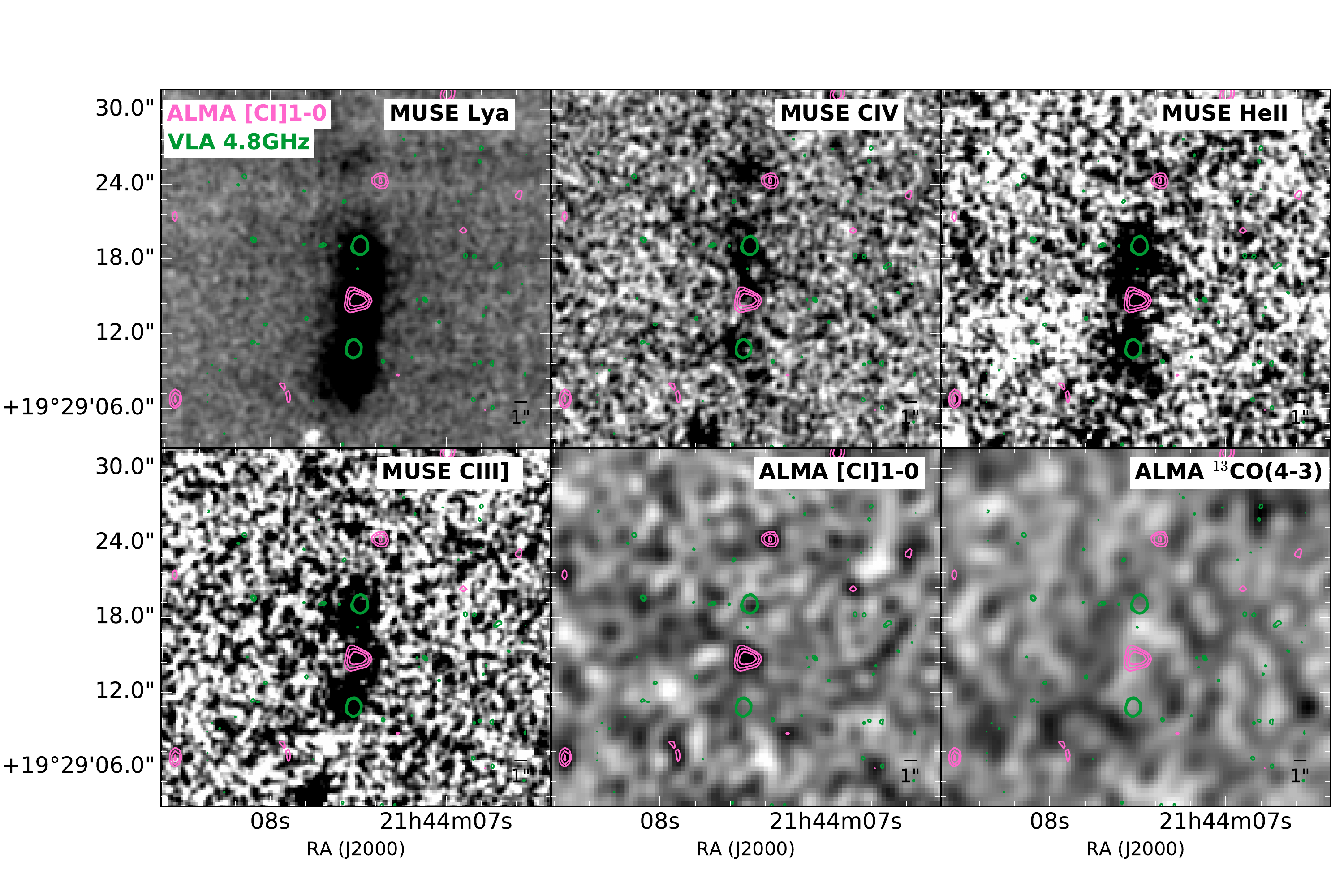}
\caption{Six emission line images constructed using the MUSE and ALMA observations of \galaxy\ (see text for details). \textit{Top left:} \Lya\ (extracted from $\lambda_{obs}$=5581.64--5584.14\,\AA); \textit{top center:} \CIV\ 1548.2\,\AA\,(extracted from $\lambda_{obs}$=7100.39--7102.89\,\AA); \textit{top right:} \HeII\ (extracted from $\lambda_{obs}$=7525.39--7535.39\,\AA); \textit{bottom left} \CIII\ (extracted from $\lambda_{obs}$=8739.14--8756.64\,\AA); \textit{bottom center:} \CIlower\ (extracted from 107.21--107.27\,GHz) from the continuum subtracted cube; \textit{bottom right:} \CO13\ (extracted from 96.007--96.067\,GHz) from continuum subtracted cube. MUSE moment-0 maps are smoothed with a Gaussian filter of size 7$\times$7 pixels. Pink contours are overlaid \CIlower\ line emission (levels at $2.5\sigma$, $3\sigma,$ and $3.5\sigma$, where $\sigma=29$\,mJy). Green contours represent the \vla\ band C data with levels at $3\sigma$, $\sqrt{2}\times3\sigma$, $3\sqrt{2}\times3\sigma,$ and $5\sqrt{2}\times3\sigma,$ where $\sigma=45$\,mJy. Since we did not extract the full \Lya\ line profile making the map due to the impact of the bright 5755\AA\ night sky line on the \Lya\ line profile, we are likely missing a significant amount of the flux and have an incomplete map of its flux distribution.}
\label{fig:Narrow_band}
\end{figure*}

\begin{figure}[ht]
\includegraphics[width=\linewidth]{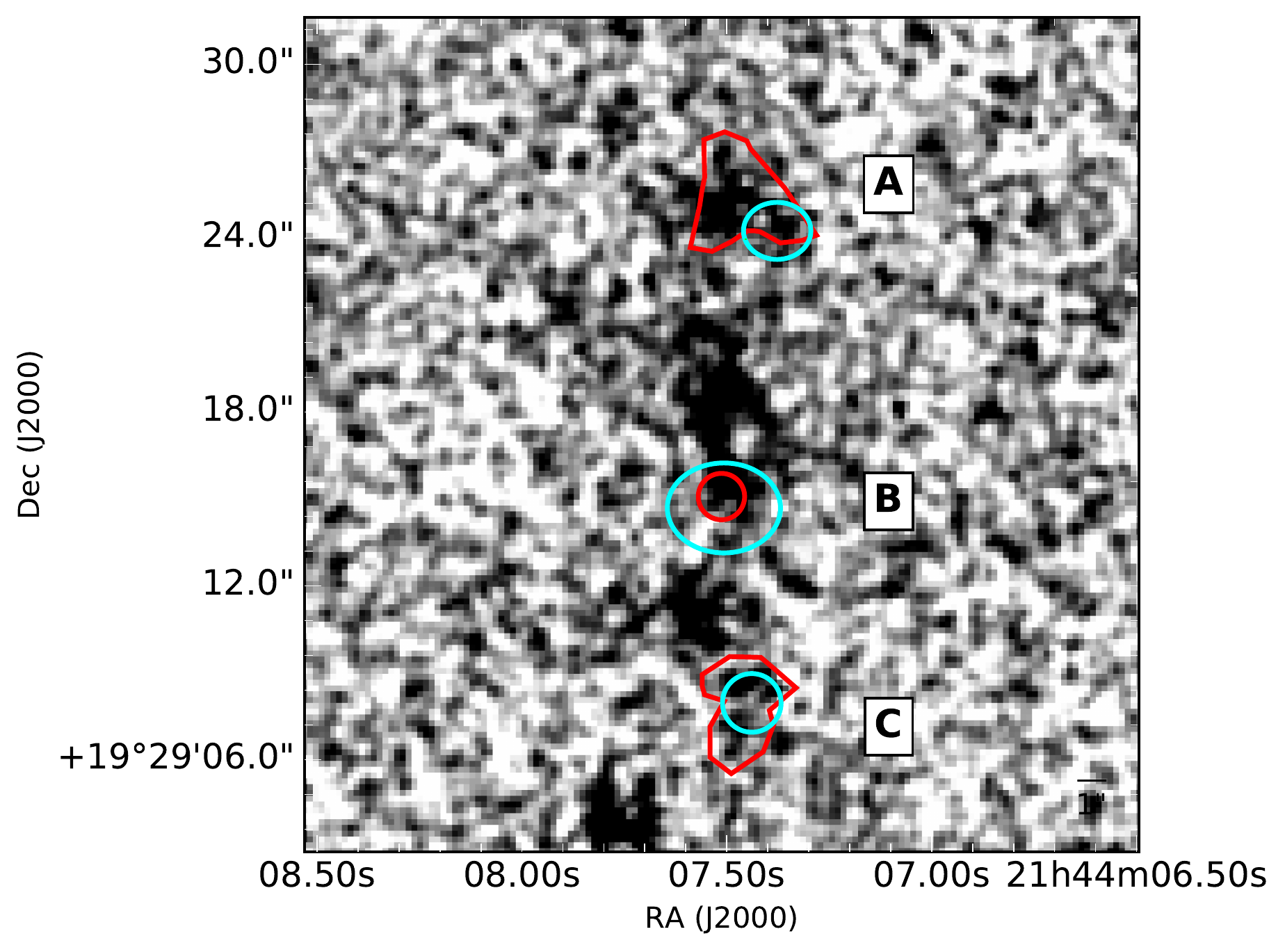}
\caption{Gray-scale images of \CIV\ emission in \galaxy. The regions bound by red lines are positions of the spectra extracted from the MUSE cube, determined from the \CIV\ narrow-band image extracted from the MUSE data cube. The regions bound by cyan lines indicate the positions of the extracted \CI\ spectra from the ALMA cube, defined from the \CI\ moment-0 map to maximize the resulting signal-to-noise. The northern \CI\ detection is unresolved; we extracted the spectrum over one beam area. The extracted spectra are shown in Fig.~\ref{fig:MUSE_ALMA_lines}. The three main regions where the spectra are extracted are indicated by an `A' for the region north of the northern radio lobe, a `B' for the host galaxy and core where the \agn\ resides, and a `C' for the region south of the southern radio lobe. The regions of \CIV\ (regions bounded by red lines) and \CI\ (bounded by cyan lines) are not perfectly co-spatial in regions A and C but the \CIV\ emission in these regions encompasses the corresponding regions of \CI\ emission.}
\label{fig:CIV_with_extracted_regions}
\end{figure}

The MUSE spectrum shows an extended structure of ionized gas around \galaxy. Qualitatively, the spectrum is consistent with the previous long-slit spectroscopy observations \citep{Maxfield2002} but now with full spatial information over the large 1$\times$1\,arcmin field-of-view of MUSE. What is observed with MUSE and not seen in previous studies are narrow emission lines of very faint extended gas - extending far beyond the radio lobes (Fig.~\ref{fig:Narrow_band}) - and the detection of regions A and C in \CIV\ (Fig.~\ref{fig:CIV_with_extracted_regions}). Over the area centered on the host galaxy, region B, strong \Lya, \CIV,\ and \HeII\ emission is detected and the extracted spectra indicate that at least one strong \Lya\ and \CIV\ absorber is present. Furthermore, the \Lya\ and \HeII\ are marginally detected in the extended regions A and C. The low robustness of these detections are due to both the intrinsic faintness of the gas and the fact that \galaxy\ is at a very unfortunate redshift. We use the term unfortunate because both the \Lya\ and \HeII\ spectra are severely affected by strong night sky lines. Due to the difficulty in subtracting night sky lines, which results in strong residuals and the additional noise they add to the spectra, it is difficult to accurately estimate the line parameters throughout the cube of these two emission lines. The very strong \OI\ night sky line at $\lambda_{\rm rest}$=5577.338\,\AA\ falls at almost the exact same wavelength as the redshifted \Lya\ emission. The \HeII\ emission is affected by three different (weaker) sky lines, which also adversely affect our ability to accurately determine and subtract the underlying continuum emission. This means that the relatively low intrinsic flux from regions A and C in both these lines is uncertain; it also means that it is difficult to fit a line profile to the emission from region B, even though it is much brighter than the emission from A or C. The emission from the non-resonant \HeII\ line is often used to determine the systemic redshift of radio galaxies and can act as a template to determine the associated absorption line profiles of the \Lya\ and \CIV\ emission lines under the assumption that all lines originate from the same gas. In the case of \galaxy,  we instead relied on the \CIlower\ line to determine the systemic redshift, and it is beyond the scope of this paper to characterize the absorbers in region B.

\CIV\ is not affected by sky line emission and this means we can fit the line as observed. Thus we can compare the \CIV\ emission with the \CI\ line detected with ALMA. This is something new and important for modeling since it provides information about the same species in different ionization states and in a different gas phase. Furthermore, even though the parameters of the \Lya\ and \HeII\ emission lines are difficult to estimate accurately, we can still determine flux ratios between different lines and recognize that they are relatively uncertain. This is important for photoionization modeling, where flux ratios between different atoms and ionization states provide direct constraints on the physical state of the gas. 

\begin{figure*}[ht]
\minipage{0.33\textwidth}
  \includegraphics[width=\linewidth]{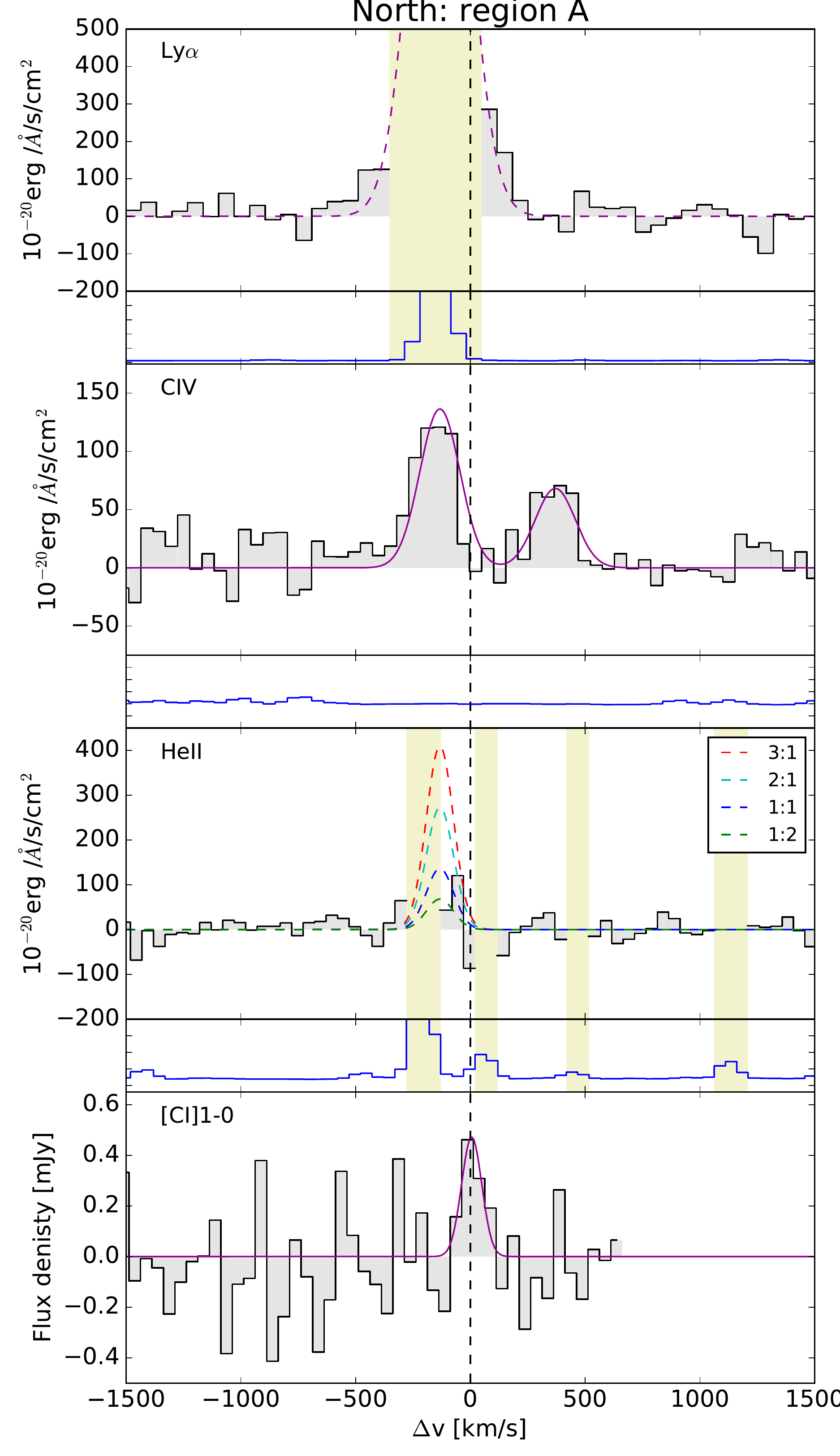}
\endminipage\hfill
\minipage{0.33\textwidth}
  \includegraphics[width=\linewidth]{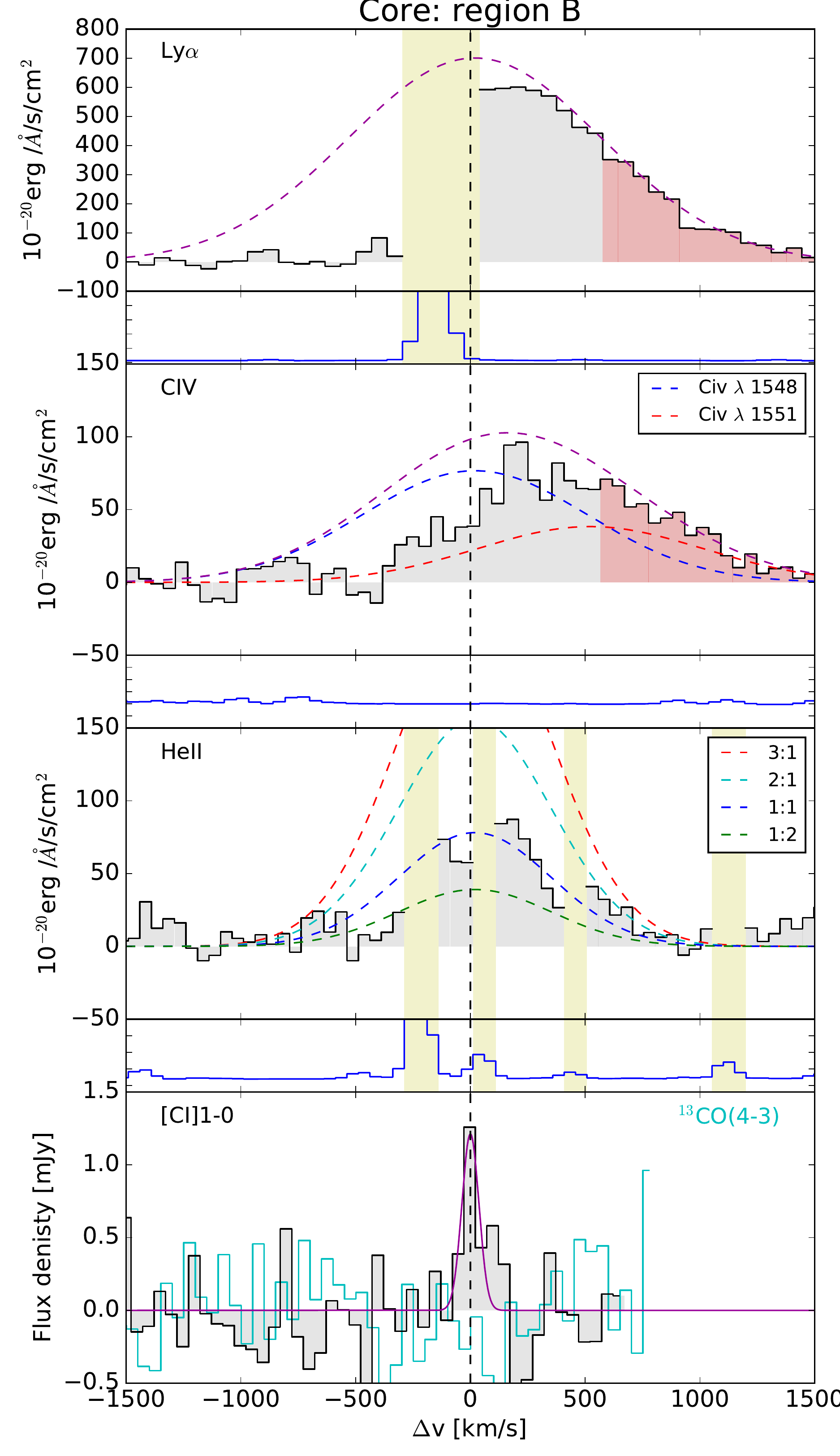}
\endminipage\hfill
\minipage{0.33\textwidth}%
  \includegraphics[width=\linewidth]{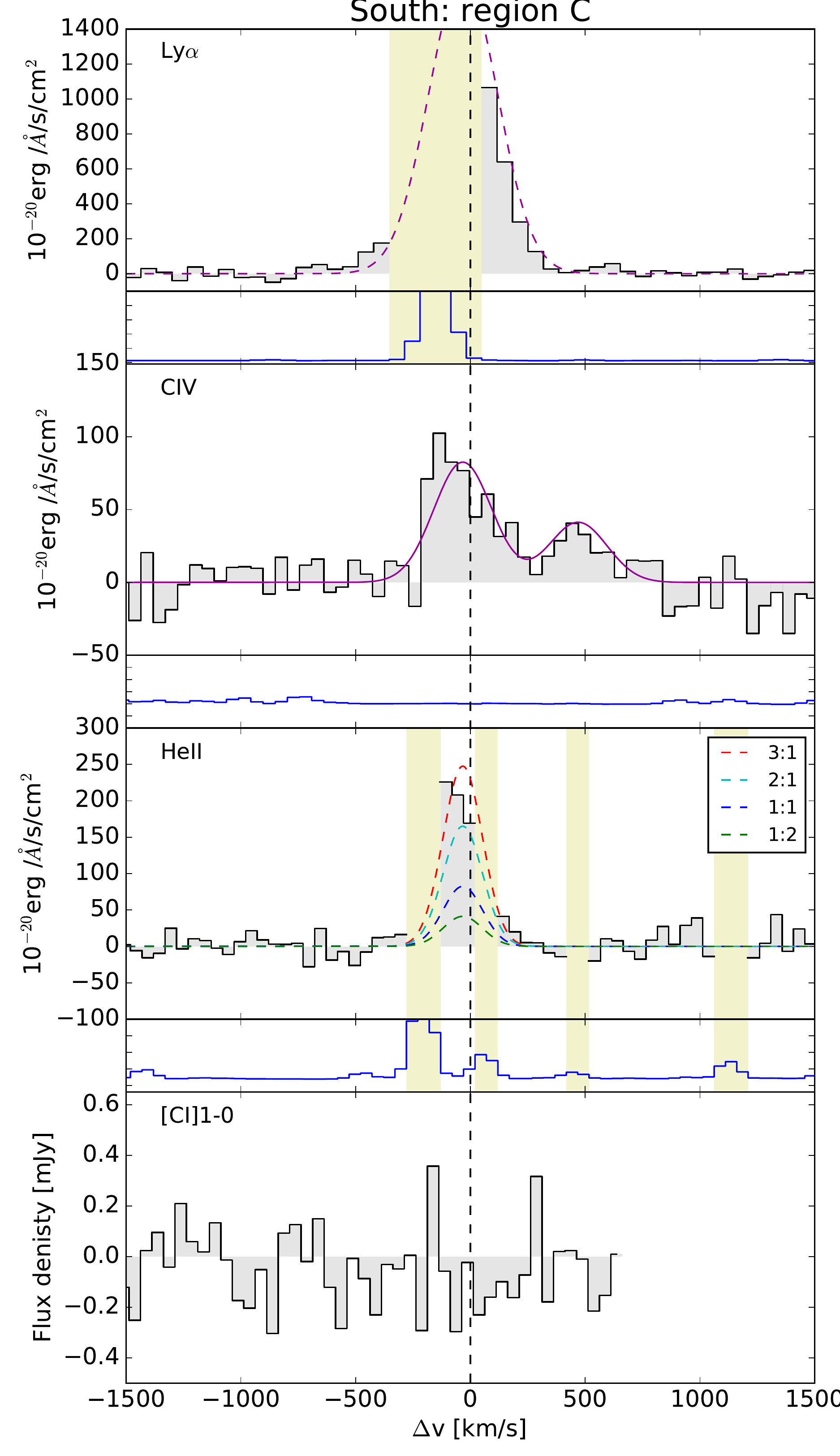}
\endminipage
\caption{Extracted spectra of three different regions, as indicated in Fig.~\ref{fig:CIV_with_extracted_regions}. \textit{Left to right:} Northern, core, and southern regions of the same emission line (which is indicated in the top left corner of each panel). \textit{Top to bottom:} \Lya, \CIV, and \HeII\ from the MUSE cube and \CI\ from the ALMA cube for each of the three regions.
The abscissa in each panel is the velocity relative to the
systemic redshift in units of \kms, estimated using the redshift of the \CI\ line at the core; the dashed vertical black line indicates the systemic velocity, which is set to zero.
The ordinate in each panel is the flux density in units of erg s$^{-1}$ cm$^{-2}$\,\AA$^{-1}$. In the panels that show the extracted spectra around the wavelengths of the \Lya, \CIV, and \HeII\ emission lines, we indicate velocity (wavelength) regions that are strongly affected night sky lines with yellow shading, and these regions are completely ignored for any fits to the line profiles.
The purple lines show the best fitting line profiles, while the dashed lines indicate those fits where a significant part of the line has been ignored in constraining the fit (Sects.~\ref{sec:alma_lines} and \ref{sec:muse_lines}). The red shaded part of the \Lya\ and \CIV\ spectrum of region B indicates the part that was used for the line fitting, and the dashed blue and red lines indicate the individual part of the \CIV\ doublet. Most of the flux in regions A and C is lost due to the impact of the strong night sky line at 5577\,\AA. Dashed blue, green, cyan, and red lines indicate possible \HeII\ lines profiles, assuming that the relative velocity and \fwhm\ are the same as that for the \CIV\ line 
and have flux ratios \HeII/\CIV\ of 0.5, 1, 2, and 3. The solid cyan curve in the lowest middle panel indicates the spectrum of the \CO13\ line around the systemic velocity (it is a non-detection and this is simply a spectrum of the noise).}
\label{fig:MUSE_ALMA_lines}
\end{figure*}

The \CIV\,$\lambda\lambda$1548,1551 doublet shows narrow line widths in the regions $\gtrsim$10\,kpc from the \agn. Both regions A and C have \CIV\ emission with sufficiently narrow lines to clearly resolve the doublet. The \CIV\ doublets are fitted with a simple double Gaussian profile to estimate the integrated line flux, line width, and velocity offset. We used a non-linear least square method to fit a double Gaussian function to the spectrum using the extracted variance spectrum to estimate the flux uncertainty. The reported uncertainties are the square root of the variance, one sigma, of the parameter estimate for the Gaussian fit. For the double Gaussian, we fixed the \CIV\,$\lambda$1548/\CIV\,$\lambda$1550 doublet ratio to the theoretical ratio of 2:1 \citep[e.g.,][]{Flower1979,Nussbaumer1981}, the width of the two lines are set to be the same, and the center of the blue component ($\lambda$1548.2\,\AA) is a free parameter with the center of the red component ($\lambda$1550.8\,\AA) fixed to a 2.6\,\AA\ shift with respect to the blue line. The best fit for the extracted spectrum of region A yields an \fwhm\ of 179$\pm$27\,\kms, blueshifted by $\sim$\,$-$133\,\kms\ (centered on the blue component) from the systemic redshift, and the doublet has an integrated flux of (11.00$\pm$2.10)$\times$10$^{-18}$\,erg\,cm$^{-2}$\,s$^{-1}$. The extracted spectra from region C show a similar line profile, with an \fwhm\ of 264$\pm$48\,\kms, blueshifted $\sim-$35\,\kms, and with an integrated flux density of (8.70$\pm$2.16)$\times$ 10$^{-18}$\,erg\,cm$^{-2}$\,s$^{-1}$. Figure~\ref{fig:MUSE_ALMA_lines} shows the best fit for regions A and B and Table~\ref{table:Line_fluxes} enumerates the line properties.

To estimate the \Lya\ $\lambda$1215.7\,\AA\ integrated line flux, we used two independent methods. First, we integrated the velocity bins not affected by sky line subtraction residuals. For regions A and C, the blue and red wings of the emission lines are present in the spectrum, while for region B only the red wing is detected. We integrated the flux over the channels containing the blue and red wings and this yields a lower limit for the integrated \Lya\ flux. Our second approach was to use the information from the other lines (\CIV\ and \CI) to predict the redshift of the center of the \Lya\ emission, and fit a single Gaussian emission profile only to the parts of the \Lya\ line that are not affected by sky lines or associated absorbers. This technique is rather straightforward for regions A and C, as the narrow \CIV\ is clearly detected; this provides  consistent results, leaving only the \Lya\ peak as an unconstrained  parameter while letting the \fwhm\ vary within the uncertainties of the \CIV\ line. Figure~\ref{fig:MUSE_ALMA_lines} shows this single Gaussian fit as a dashed purple line.

This method is more complicated for region B due to the presence of a strong associated absorber seen in both \Lya\ and \CIV. Such absorbers have been observed in several other \hzrgs\  \citep[e.g.,][]{kolwa2019}. An additional complication is the overlap of the two \CIV\ doublet lines since both have associated absorption lines. For this region, we therefore started by fitting the red wing of \Lya\ at velocities $>$600\,km\,s$^{-1}$. This avoids the peak region where the associated absorber "spills over" to the red side. We then used the \Lya\ results to fit the \CIV\ line with a fixed 2:1 ratio for the two doublet lines, while keeping the redshift fixed to the \Lya\ and \CI\ redshift. Figure \ref{fig:MUSE_ALMA_lines} shows this double Gaussian fit as a dashed purple line, with the individual lines also shown in blue and red. In a forthcoming paper, we plan to fit these lines with consistent Voigt profiles to determine the column densities in the associated absorbers. We consider the "maximum" \Lya\ flux derived from our Gaussian fit as the most reliable measurement. We will only use those values in the remainder of the paper.

The \HeII\ $\lambda 1640.4$\,\AA\ emission is significantly harder to fit due to the presence of three sky lines and the uncertainty in the continuum subtraction. We therefore estimated the flux using a fully constrained (i.e., completely scaled) Gaussian by assuming both the redshift and \fwhm\ from the \CIV\ lines in each region. We plotted the corresponding scaled line profiles for \HeII/\CIV\ line flux ratios 1:2, 1:1, 2:1, and 3:1 (Fig.~\ref{fig:MUSE_ALMA_lines}). If we can only trust the channels unaffected by strong sky lines, then this tells us that the \HeII\ spectrum of region A is consistent with a ratio of 1:2 to 2:1, close to 1:1 in the \agn\ region B, and between 2:1 and 3:1 in region C. The \HeII\ is apparently only brighter than \CIV\  in region C.

\begin{table*} 
\centering
\begin{tabular}{l l l}
\toprule
\toprule
Property & Value & Reference\\
\midrule
Ra(J2000), Dec (J2000)                          &21:44:7.45, +19:29:14.60&                      \\
Ra($^\circ$), Dec($^\circ$)                     &326.03104, 19.48739    \\
Systemic redshift ($z_{\rm sys}$)       & 3.5895                                        & This work \S~\ref{sec:alma_lines}\\
Stellar mass (\mstar)                           & $10^{11.13}$M$_{\odot}$       & \cite{DeBreuck2010}\\
Molecular mass (\mmol)          &$3.1 \pm 1.1 \times 10^{10}$\,M$_\odot$                                                        &This work \S~\ref{sec:molecular_gas_mass_estimate}\\
Ionized mass (\mion)            &$1--18\times 10^7$\,\msun      &This work \S~\ref{sec:molecular_gas_mass_estimate}\\
\vspace{0.5mm}
Infrared luminosity (L$_{\rm IR} \equiv L_{8-1000\mu {\rm m}}$) & 0.74$_{-0.55}^{+1.52}\times 10^{12}$\,L$_{\odot}$  &\cite{Falkendal2019}\\
\vspace{0.5mm}
Star-formation rate (\sfr)                      & 84$_{-62}^{+172}$\,M$_{\odot}$\,yr$^{-1}$     & \cite{Falkendal2019}\\
\vspace{0.5mm}
Star-formation efficiency (SFE)         & 2.74$^{+5.57}_{-0.99}$\,Gyr$^{-1}$                            & This work \S~\ref{sec:molecular_gas_mass_estimate}\\
Depletion time scale ($\tau_{\rm dep}$) &       0.36$^{+1.08}_{-0.18}$\,Gyr                                     & This work \S~\ref{sec:molecular_gas_mass_estimate}\\
Gas fraction (\fgas)    &       0.19$\pm$ 0.07  & This work \S~\ref{sec:discussion_host_galaxy}\\
\sfr\ surface density (log$\Sigma_{\rm SFR}$)   & -0.87\,\msun\,kpc$^{-1}$      & This work \S~\ref{sec:discussion_host_galaxy}\\
Gas surface density (log$\Sigma_{\rm gas}$)     & 1.69\,\msun\,pc$^{-1}$        & This work \S~\ref{sec:discussion_host_galaxy}\\
\bottomrule
\end{tabular}
\caption{Properties of the host galaxy 4C~19.71 (MG~2144+1928). The systemic redshift is estimated from the redshift of the \CI\ line emission at the location of the host galaxy and \agn\ (region B). The \sfr\ is estimated using the infrared luminosity, L$_{\rm IR}$, and scaling it using the conversion given in \cite{Kennicutt1998}, but for a Kroupa initial mass function (\imf) (we divided the original conversion factor by 1.5 since the \sfr\ is calculated using a Salpeter \imf\ in \citealt{DeBreuck2010}).}
\label{table:prop}  
\end{table*}

\begin{table*}      
\centering
\begin{adjustbox}{width=1\textwidth}          
\begin{tabular}{+l ^c ^c ^c ^c ^c ^c ^c }
\toprule
\midrule
\multicolumn{8}{c}{ALMA} \\
\midrule
 Line & $\nu_{\text{rest}}$ & $\nu_{\text{obs}}$ & $\Delta$v & peak flux & $S$d$V$ & Line flux &  \fwhm \\ 
 & GHz   & GHz    & \kms  & mJy & Jy\,\kms &$10^{-18}$\,erg\,cm$^{-2}$\,s$^{-1}$   & \kms  \\
\midrule
\multicolumn{5}{l}{\textbf{North (region A)}} \\
\CI & 492.2             &107.24 &5$\pm$23               &0.47$\pm$0.20  &$\la$0.05$\pm$0.03     &0.19$\pm$0.12  &108$\pm$54\\ 
\multicolumn{5}{l}{\textbf{Core (region B)}} \\
\CI & 492.2     &107.23 &0                      &1.21$\pm$0.29  &0.11$\pm$0.04         &0.40$\pm$0.15  &87$\pm$23\\
\multicolumn{5}{l}{\textbf{South (region C)}} \\
\CI & 492.2     &107.23 &  \nodata      & \nodata       & <0.06 & \nodata       & \nodata\\
\midrule
\multicolumn{8}{c}{MUSE} \\
\midrule
Line & $\lambda_{\text{\rm rest}}$ & $\lambda_{\text{\rm obs}}$ &$\Delta$v & peak flux  &  Line flux  & Line flux & \fwhm \\
        & \AA  & \AA & \kms  & $10^{-20}$\,erg\,cm$^{-2}$\,s$^{-1}$\,\AA$^{-1}$ &$10^{-16}$\,erg\,cm$^{-2}$\,s$^{-1}$\,\AA$^{-1}$\,\kms  & $10^{-18}$\,erg\,cm$^{-2}$\,s$^{-1}$ &    \kms  \\
\midrule
\multicolumn{5}{l}{\textbf{North (region A)}} \\
\Lya\ (min)$^*$ &1215.7                 & \nodata       & \nodata               &$>$280                 &$>$5.8                 &$>$13.7                &$\sim$179       \\
\Lya\ (max)$^\dagger$   &1215.7                 &fixed to \CIV          &fixed to \CIV         & $\sim$1500                    &$\sim$3450                     &$\sim$94               &$\sim$230      \\
\CIV        &1548.2 (1550.8)    &7102.3$\pm$0.3 &-133$\pm$14    &138$\pm$18         &2.5$\pm$0.7 &11.0$\pm$2.1   &179$\pm$27\\
\HeII$^{\star\star}$    &1640.4                 &fixed to \CIV  &fixed to \CIV    &70--276                &1.3--5         &5.5--22                &fixed to \CIV \\
\multicolumn{5}{l}{\textbf{Core (region B)}} \\
\Lya\ (min)$^\star$     &1215.7                 &\nodata\       &\nodata\               &$>$600                 &$>$44.0                &$>$104         &$\gtrsim$1000\\
\Lya\ (max)$^\dagger$   &1215.7                 &5580.4$\pm$0.3 &50$\pm$16              &700.9$\pm$19                   &90.6$\pm$18.5 & 179$\pm$5     &1300$\pm$230 \\
\CIV    &1548.2 (1550.8)   &fixed to \Lya  &fixed to \Lya               &102.8$\pm$0.6                  & 11.8$\pm$0.9 & 33.5$\pm$0.2     &1148$\pm$12 \\
\HeII$^{\star\star}$  &1640.4  &fixed to \CIV  &fixed to \CIV           &51--102                        &5.9--11.8 &16.7--33.5 &fixed to \CIV \\
\multicolumn{5}{l}{\textbf{South (region C)}} \\
\Lya\ (min)$^*$ & 1215.7                & \nodata       & \nodata       &$>$1000                &$>$26.6                &$>$63          &$\sim$264\\
\Lya\ (max)$^\dagger$   &1215.7                 &fixed to \CIV  &fixed to \CIV            &$\sim$1700             &$\sim$230              &$\sim$130      &$\sim$320\\
\CIV  &1548.2 (1550.8)  &7104.7$\pm$0.6         &-35$\pm$25     &80$\pm$14      &2.1$\pm$0.8    &8.70$\pm$2.16  &264$\pm$48\\
\HeII$^{\star\star}$    &1640.4                 &fixed to \CIV  &fixed to \CIV            &160--240               &7.3--11.0              &17.4--26.1             &$\sim$264      \\
\bottomrule
\end{tabular}
\end{adjustbox}
\caption{\small  Observed line fluxes and widths in the three regions, both in velocity integrated and frequency- (ALMA line) or wavelength-integrated (MUSE). For the MUSE lines, we give the \fwhm,\ which has been corrected in quadrature using the instrument resolution of $\sim$110\,\kms\ at $\sim$1550\,\AA. In the southern region, the \CI\ flux is a 3$\sigma$ upper limit.
($^\star$) The minimum \Lya\ flux is estimated by integrating over the wavelength range where the emission should be, excluding the wavelength ranges affected by sky lines (yellow regions in Fig.~\ref{fig:MUSE_ALMA_lines}). At the core of \galaxy, the blue side is completely absorbed and dominated by sky-subtraction residuals from the strong 5577\AA\,sky line; the integrated flux is thus a lower limit since we are missing a significant portion of the likely emission.
($^{\dagger}$) Fit to the \Lya\ line excluding the wavelengths impacted by the night sky residuals. These fits are shown as the dashed red line in Fig.~\ref{fig:MUSE_ALMA_lines}.
We note that these estimates are likely to only be accurate to the order-of-magnitude level given the limited range of the profile that is not impacted by the strong night sky line at these wavelengths. ($^{\star\star}$) The \HeII\ line is severely affected by sky line noise and residuals and cannot be fitted with any certainty. Instead, the integrated flux is estimated by scaling a Gaussian with the same width and velocity offset as the \CIV\ line, and for flux ratios \HeII/\CIV\ = 0.5, 1, and 2 for the northern component and \HeII/\CIV\ = 2 and 3 for the southern component, as indicated in Fig.~\ref{fig:MUSE_ALMA_lines}.} 
\label{table:Line_fluxes}
\end{table*}

\subsection{Foreground objects}
\label{sec:foreground_objects}

In the field around \galaxy, four foreground galaxies are detected (Fig.~\ref{fig:support_data}). Galaxy 1 is located south of the southern radio lobe at z=0.483 and detected in X-ray, K-band, and IRAC 1. Galaxy 2 at z=3.31 is also located in the southern part of the field; however, it is not seen in any continuum and was detected in line emission while searching through the MUSE cube. Galaxy 3 is located close to the hotspot of the southern radio lobe and is at z=0.693; it was detected in MUSE but not seen in any continuum emission. Galaxy 4 is detected in the K-band and IRAC 1 image and is at z=1.03, and not at the redshift of \galaxy\ as previously reported from long-slit spectroscopy of \Lya\ \citep{Maxfield2002}. In the aperture of Galaxy 4, \Lya\ is detected at the same redshift as \galaxy\ in the extracted MUSE spectra, but this is just weak extended emission from \galaxy\ itself and not \Lya\ emission from Galaxy 4. More details about the redshift confirmation of the four foreground galaxies are available in Appendix~\ref{sec:redshift_of_foreground}.

\section{Discussion}

\subsection{A normal star-forming galaxy?}
\label{sec:discussion_host_galaxy}

Galaxy \galaxy\ is a massive high-z radio galaxy with a relatively low SFR of $\sim$90\,\msun\,yr$^{-1}$ and it falls below the main sequence of star-forming galaxies \citep{Falkendal2019}. The star formation efficiency, SFE=2.74\,Gyr$^{-1}$, or gas depletion time, $\tau_{\rm dep}$=0.36\,Gyr, are somewhat lower than, but are consistent overall with, star-forming galaxies at high-$z$ \citep{Daddi2010gas_fraction,Tacconi2013,Elbaz2017}. We estimated a gas fraction, \fgas = \mmol/(\mmol+\mstar), of \fgas=0.19$\pm 0.07$; this is low in comparison to star-forming galaxies at $z\sim$1-3, which have an average \fgas$\sim$0.5 \citep{Daddi2010gas_fraction,Tacconi2013,Elbaz2017}. All of these values, the short gas depletion time and low gas fraction, are typical of other HzRGs and some submm galaxies \citep{man19}. \cite{Santini2014} find that the gas fraction increases with \sfr\ and decreases with stellar mass for main sequence galaxies out to $z\sim$2.5. For galaxies with a constant log (\mstar/\msun)$\approx$11.5 at $z\sim$2.5, they find a gas fraction of \fgas$\sim$0.2; for the same \mstar\ but at a fixed log SFR$\sim$2, they find \fgas$\sim$0.5. This is consistent with our estimated gas fraction, since \galaxy\ has a high stellar mass log(\mstar/\msun)=11.13\,\msun\ \citep{DeBreuck2010} and low \sfr. It should be emphasized that the work by \cite{Santini2014} only includes normal star-forming galaxies, not AGNs, and does not cover the redshift of \galaxy. To characterize \galaxy\ further, we estimated the SFR surface density, $\Sigma_{\rm SFR}$, and the gas surface density, $\Sigma_{\rm gas}$, by assuming the size of an ALMA beam (1.9\arcsec$\times$1.8\arcsec) as the size of both the gas and stellar component \citep[which has been shown to be the case for other radio galaxies, e.g.,][]{Miley1992, Pentericci2001, Emonts2015}. We estimated log$\Sigma_{\rm SFR}$=-0.87\,\msun\,kpc$^{-2}$ and log$\Sigma_{\rm gas}$=1.69\,\msun\,pc$^{-2}$, both of which are consistent with  \galaxy\ lying along the Schmidt-Kennicutt relation within the large uncertainty of our estimated \sfr\ and the scatter of the relationship itself \citep{Daddi2010K-S-relationship,Kennicutt1998K-S_relationship,Genzel2010}. Thus \galaxy\ appears to be a normal star-forming galaxy with a relatively low SFR.

\subsection{Galaxy disk misaligned with nuclear launching region?}
\label{sec:diffuse_gas}

The kinematics of the submm \CI\ and the rest-frame UV line \Lya\ line emission from the nucleus of \galaxy\ are vastly different, with \fwhm s\ of $\sim$90\,\kms\ and $\gtrsim$1000\,\kms, respectively. The width of the \Lya\ emission is only a crude estimate, and we do not know if the broadening is due to resonance scattering, scattered broad line emission, or the intrinsic kinematics of the warm ionized gas around the \agn\ and circum-nuclear region. Still, it is clear that the \Lya\ is much broader than the \CI\ over this region. The \CI\ line is very narrow and dynamically cold. The gas traced by \CI\ cannot be within the gravitational influence of the \agn. The gas is not part of an outflow. It must be outside the ionization cone of the \agn\  because, unlike other emission lines from the nuclear regions, it is much more dynamically quiescent. The core of the radio source is not detected (Fig.~\ref{fig:support_data}) and the source is lobe dominated. The core to lobe fraction can serve as an orientation indicator \citep{Kapahi1982,Drouart2012}. We are likely observing the radio jets propagating more or less in the plane of the sky.

If the \CI\ gas is confined within the disk of the host galaxy, we would expect the \CI\ line width to be larger, unless we are observing the disk almost face-on. To confirm this, we estimated the dynamical mass \citep[following][]{Feng2014}, taking the radius of the galaxy as half a beam size assuming the \CI\ line probes the total stellar plus gas mass and that it is equal to the dynamical mass. This results in an inclination of $i\sim 3^{\circ}$ and indicates that we are viewing a rotating disk traced by \CI\ emission almost face-on. This means that the galaxy is rotating in the plane of the sky and the radio structure is also propagating in the plane of the sky. Since it is thought that the radio jets are propagating perpendicular to the accretion disk, this means that the orientation of the nuclear launch region is not aligned with the host galaxy. This is not impossible and has been observed in the nearby universe \citep[e.g.,][]{Morganti1998}. This is also not the first \hzrg\  showing narrow molecular line emission at the nucleus. \citet{Gullberg2016_0943} found molecular gas traced by CO(8--7) with \fwhm=43$\pm$13\,\kms\ at the nucleus for the high-$z$ radio galaxy MRC~0943-242. The galaxy has extended \Lya\ emission with \fwhm=1592$\pm$44\,\kms\ within the circum-nuclear region. This galaxy shows similar properties to what we observed and could also be explained as being viewed face-on and having a misalignment between the central nuclear launch region and the galaxy disk.

\begin{figure*}[!ht]
\includegraphics[width=\linewidth]{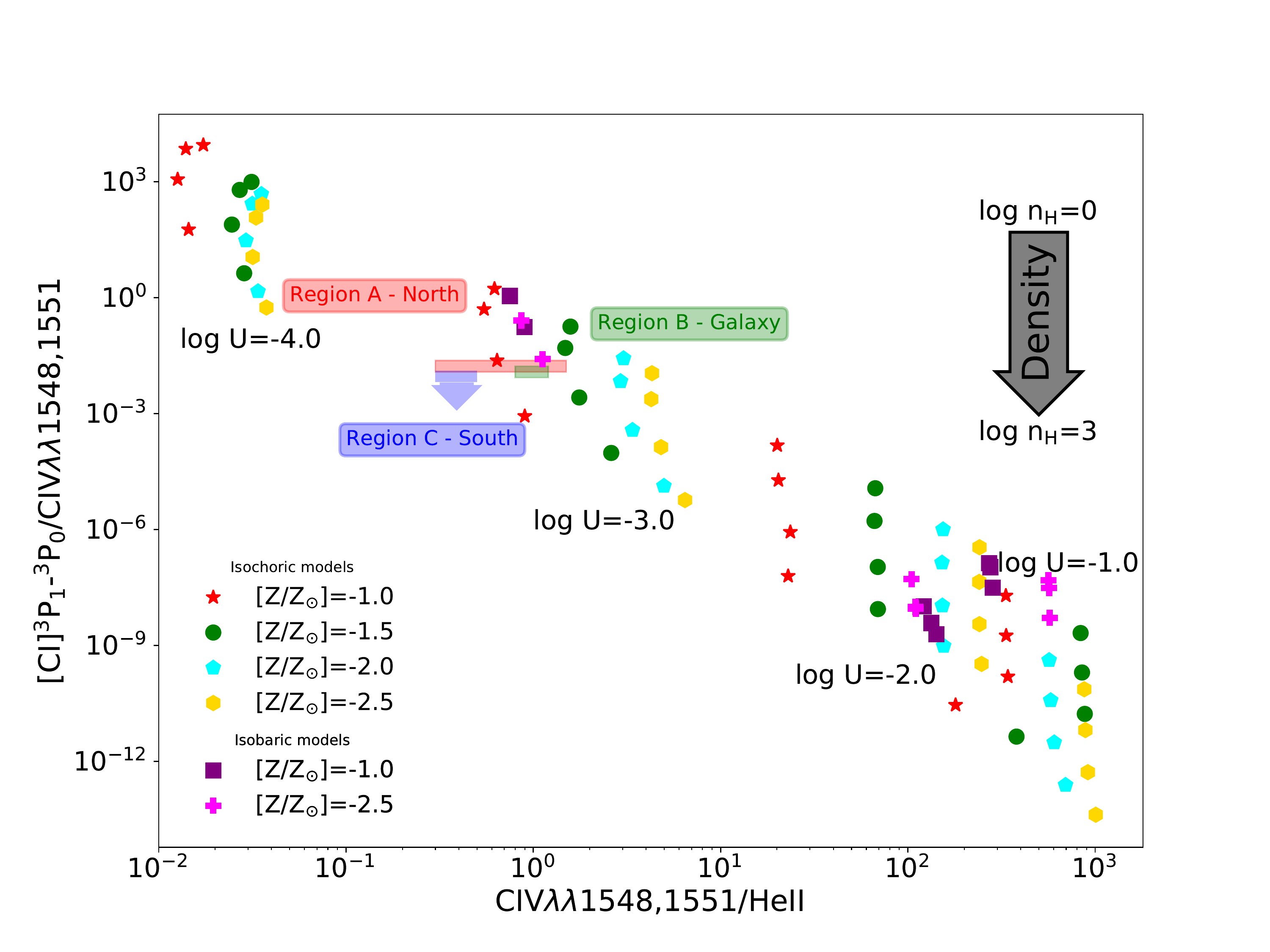}
\caption{The line ratios of \CI/\CIV\ as a function of \CIV/\HeII\ for the three regions as defined in Fig.~\ref{fig:CIV_with_extracted_regions} and Cloudy models. The colored points represent the photoionization and photodissociated region (\pdr) models as indicated in the legend. The logarithm of the ionization parameter of each of the groups of points in the figure
are indicated in the appropriate regions (decreasing from $-$1 to $-$4\,dex from
left to right).  For each of the isochoric models, the density of the
gas increases upward from log n$_H$ (\cmthree)=0 to 3. We also show a set of isobaric
models for three pressures, log P/k (K\,\cmthree)=2, 3, and 4. The three colored boxes show the estimated line ratios, as estimated from our observations (red for region A; green for region B; and blue, which is an upper limit in \CI/\CIV,\ for region C). We note that the box shown exaggerates the uncertainty in \CI/\CIV\ and that it shows the range of plausible values for \CIV/\HeII\ and not the uncertainties in the estimate.The apertures over which the lines fluxes were extracted are shown in Fig.~\ref{fig:CIV_with_extracted_regions}.
\label{fig:line_ratios}}
\end{figure*}

\subsection{Molecular gas mass and star formation efficiency in the radio galaxy}

We determined the molecular gas mass (\mmol) from the \CIlower\ detection at the host galaxy, following \cite{PapadopoulosGreve2004, Wagg2006, Alaghband-Zadeh2013, Gullberg2016}, as:

\begin{equation}
{\rm M}_{\rm H_2}=1375.8\frac{D^2_{\rm L}}{(1+z)} \Bigg[ \frac{X_{\rm [C\textsc{i}]}}{10^{-5}} \Bigg]^{-1} \Bigg[ \frac{A_{10}}{10^{-7}s^{-1}}\Bigg]^{-1} Q_{10}^{-1} \Bigg[ \frac{S_{\rm [C\textsc{i}]}{\rm d}V}{{\rm Jy\,km\,s^{-1}}} \Bigg]
\label{eqn:MassCI}
\end{equation}

\noindent
in units of \msun, where $D_L$ is the luminosity distance in Mpc and $X_{\rm [C\textsc{i}]}$ is the \CI-to-\Htwo\ abundance ratio; we assumed $X_{\rm [C\textsc{i}]}$=3$\times 10^{-5}$ \citep{Weiss2003}. Here, $Q_{10}$ is the excitation factor and depends on the temperature and density of the gas, as well as on the intensity of the radiation field impinging upon it \citep{Papadopoulos2004}. Without having any other molecular lines (e.g., CO or \CIupper), we cannot constrain this parameter and we assumed the median value of $Q_{10}$=0.48 \citep{PapadopoulosGreve2004, Emonts2018}. Finally, $A_{10}$ is the Einstein A-coefficient, $A_{10}$=7.93$\times$10$^{-8}$ s$^{-1}$ \citep{Papadopoulos2004}. At the location of the host galaxy, the integrated \CI\ flux density is 0.11$\pm0.04$\,Jy\,\kms,\ which results in an estimated total molecular gas mass of M$_{\rm H_2}^{\rm Core}$=(3.06$\pm$1.11)$\times$10$^{10}$\,\msun\ for the core (region B). The uncertainty in the estimated M$_{\rm H_2}$ is calculated from the one sigma error of $S_{\rm [CI]}$, without additional errors arising from from uncertainties in $X_{\rm [C\textsc{i}]}$, $Q_{10}$, or $A_{10}$. To check if our estimated value of M$_{\rm H_2}^{\rm Core}$ is reasonable, we compared it with the molecular gas mass estimated from the observed continuum flux density in ALMA band 3 \citep[via Eq.~16 in][]{Scoville2016}. For $\nu_{\rm obs}=103$\,GHz, $S_{\rm 103 GHz}=0.07$\,mJy \citep{Falkendal2019}, and assuming $T_{\rm dust}$=40\,K, we find a molecular gas mass of $M_{\rm mol}= 6 \times 10^{10}$\,\msun, which is close to the M$_{\rm H_2}^{\rm Core} \sim 3 \times$10$^{10}$\,\msun\ estimated using the \CI\ line. Considering that there are significant uncertainties in the \CI-to-\Htwo\ and CO-to-\Htwo\ conversion factors, which depend on many factors (elemental abundances, physical characteristics of the gas, radiation field impinging on the gas, cosmic ray ionization rate, etc.), and the relatively large uncertainty in the \CI\ detection, a factor of two between different estimates of the molecular gas mass is entirely reasonable and expected.

Using the molecular gas mass estimated from the \CI\ at the core and the SFR, we estimated the star formation efficiency, SFE$\equiv$SFR/\mmol=2.74$^{+5.57}_{-0.99}$\,Gyr$^{-1}$ (Table~\ref{table:prop}). The corresponding depletion time scale, $t_{\rm depl}\equiv$SFE$^{-1}$=0.36$^{+1.08}_{-0.18}$\,Gyr, is the time it takes a source to consume its molecular gas reservoir. The large uncertainty in these derived quantities arises mainly from the \sfr\ estimate. The \sfr\ is estimated from the total far-infrared (FIR) luminosity of the star-forming component of the host galaxy (integrated over $\lambda_{\rm rest}$=8--1000\,\mum) after disentangling the contribution of the warm dust emission excited by the \agn\ \citep{Falkendal2019}. The star-forming component is only constrained by two detections in the infrared (IR), which results in a large uncertainty in the \sfr\ estimate.

\subsection{The nature of the \cgm}

We observed a region $\sim$75\,kpc from the nucleus ($\sim$40\,kpc from the northern radio lobe, region A) of diffuse gas, detected in \CIV\ and \CIlower. To ensure that the \CI\ detection is not from a galaxy in the halo, we confirmed that the \CI\ is not detected in any continuum emission with ALMA, in the K-band, in any of the four IRAC bands, or with MIPS 24\,\mum. The MUSE spectrum at this location only shows weak \CIV\ (at the redshift of the radio galaxy) and does not show any additional lines, suggesting that there is no companion galaxy coinciding with the northern \CI\ detection. Four other foreground galaxies are, on the other hand, detected in the MUSE cube (see Sect.~\ref{sec:foreground_objects} and Appendix~\ref{sec:redshift_of_foreground}). 
The \CIV\ and \CI\ gas has a low velocity shift with respect to the systemic redshift of the host galaxy, $\Delta$v=5$\pm23$\,\kms\ for \CIlower\ and $\Delta$v=$-$133$\pm14$\,\kms\ for \CIV. Both lines are also very narrow: 108$\pm$54\,\kms\ and 179$\pm27$\,\kms\ for \CIlower\  and \CIV, respectively. The dynamics of the gas are quiescent compared to the gas in between the radio lobes. While the kinematics are such that it is not completely clear if the ionized gas and the molecular gas are physically related, they do overlap spatially at the relatively low spatial resolution of our data sets. 

\subsubsection{Very simple photoionization and photon-dominated region modeling} \label{subsec:photomod}

\begin{figure}[!ht]
\includegraphics[width=\linewidth]{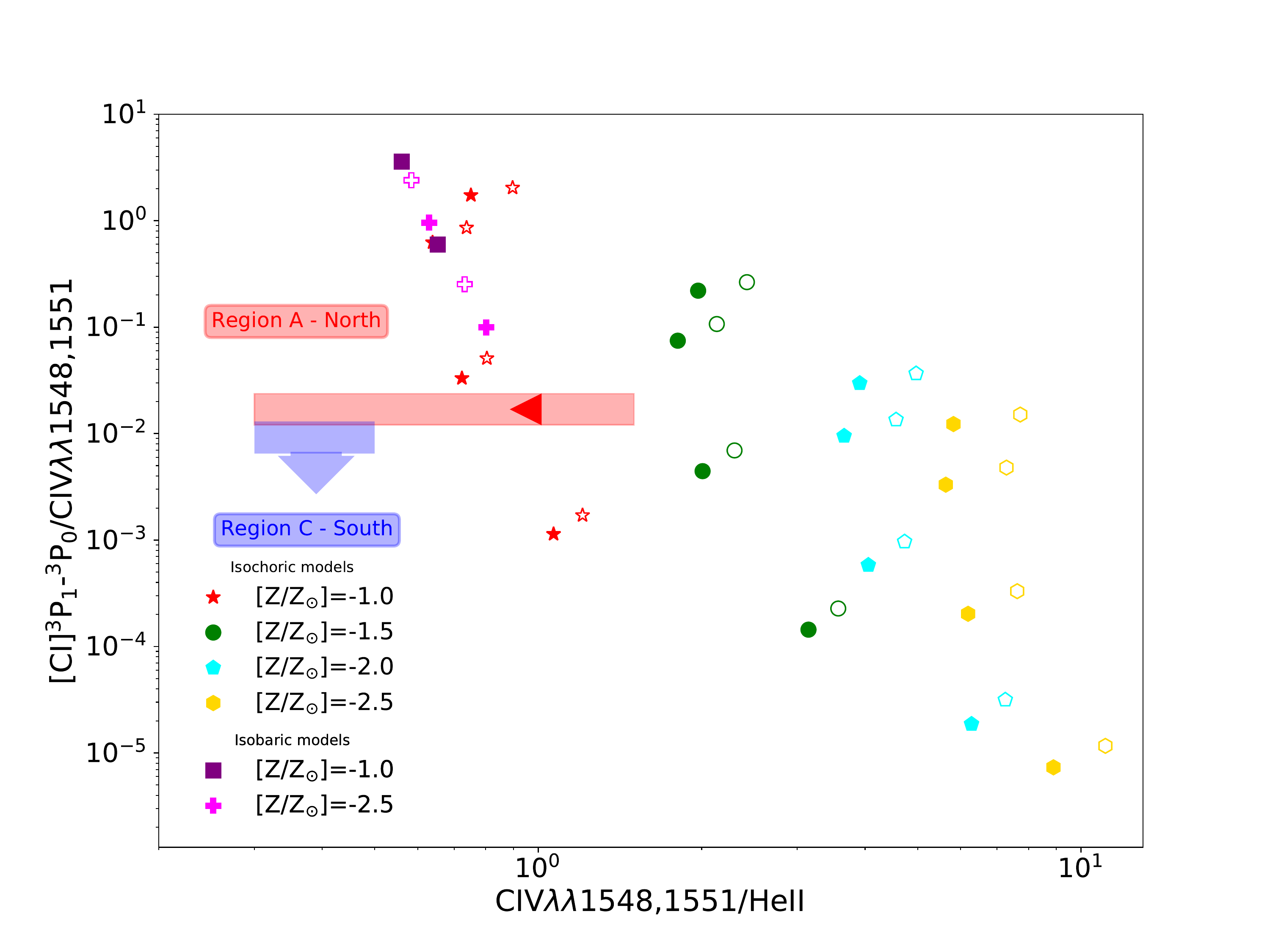}
\caption{The line ratios of \CI/\CIV\ as a function of \CIV/\HeII\ for regions A and C (using the areas shown in Fig.~\ref{fig:Narrow_band} and as labeled in the figure: red for region A and blue for region B).
The colored points represent the models as indicated in the legend and are the same as in Fig.~\ref{fig:line_ratios}. We only show the region for an ionization parameter of log U=$-$3. The solid points indicate the ratios for a model \agn\ power-law ionizing spectrum with a slope of $-$1.5, while the hollow points show the results of a model with a power-law ionizing slope of $-$2.0. This is in contrast with Fig.~\ref{fig:line_ratios}, which only shows the line ratios for a model with a power-law slope of $-$1.0. The leftward pointing arrow represents the upper limit to the ratio of \CIV/\HeII$\approx$1.
\label{fig:line_ratios_AGNslope}}
\end{figure}

In order to investigate the physical state of the multi-phase gas detected in the \cgm, we modeled the ionization front with Cloudy 17.01, most recently described by \cite{Ferland2013}. Grids of models were made assuming the standard
ISM abundance ratios scaled to a range of metallicities. As shown in Sect.~\ref{sec:Consistency_AGN_ionization_rate}, the \agn\ has sufficient luminosity to ionize the \cgm. The ionizing spectrum of the \agn\ was implemented as a simple hard power-law with a slope of $-$1 over the energy range of 0.01 to 12\,Ryd \citep[the slope is consistent with other studies of \agn s\ and the range of energies is sufficient to both excite the \pdr\ emission and to ionize He II and C IV;][]{Kraemer2000, Osterbrock2006}. We also made models with high energy limits from two to a few times higher than 12\,Ryd and found that changing this limit made no difference in the line ratios we investigated in this study. The density, for the isochoric models, varied from log n$_H$ (\cmthree)=0 to 3 in steps of 1\,dex, and the log of the ionization
parameters ranged from $-$4 to $-$1 in 1\,dex increments. Our models had an extinction of A$_{\rm V}$=2 magnitudes. In our initial exploration of models, we tried a range of extinctions, finding that increasing it beyond two did not
affect the results. For the isobaric models, we kept all the parameters the same as for the isochoric models,
now allowing the thermal pressure, log P/k (K\,\cmthree), to vary from 2 to 4 in steps of 1\,dex (we did not include any turbulent or radiation pressure). The initial densities, log n$_{\rm H}$ (\cmthree), of the isobaric models were 1.0. In order to test the sensitivity of the model results to the slope of the ionizing spectrum, we also modeled the gas illuminated by a power-law spectrum with indices of $-$1.5 and $-$2.0 \citep[see][and references therein]{lusso15}. The resulting line ratios show only modest differences compared to our fiducial model, which has a power-law spectral index of $-$1.0 (Fig.~\ref{fig:line_ratios_AGNslope}). The modeling was not intended to be exhaustive but only to determine whether or not we could reproduce the observed line ratios with a reasonable, and yet very simple, set of assumptions.

In the extended \cgm\ cloud, we measured the \CI\ and \CIV\ fluxes and could roughly estimate the \HeII\ flux. The estimates yield flux ratios of \CI/\CIV=0.017$\pm$0.011 and \CIV/\HeII$\sim$0.5-2. We note that the regions where the \CI\ and \CIV\ lines are located are not perfectly co-spatial, but the regions of \CIV\ emission encompass the \CI\ emission (upper limit) in region A (region C). This fact will be important for interpreting the nature of these regions. We show the results of the Cloudy modeling and the comparison with these line ratios in Fig.~\ref{fig:line_ratios}.
The line ratios of \CI/\CIV\ and \CIV/\HeII\ (Fig.~\ref{fig:line_ratios}) suggest that this region could be an ionization front in a molecular cloud (a photon-dominated region or a \pdr) in the \cgm,\ perhaps ionized and excited by the radiation field of the \agn\ \citep[see also][]{li19}. The characteristics of the cloud are not well constrained as a range of models provide acceptable matches with the data given both the uncertainties in the data and also the simplicity of the modeling. We can only conclude from the comparison with the photon excitation modeling that the cloud is not solar metallicity with a normal dust to gas mass ratio for the ISM of our Milky Way (not shown because they fall very low in their \CI/\CIV\ ratios) and that the gas is low metallicity for the isochoric models, about 0.1 to 0.03 solar, and has a low ionization. Regardless of the slope of the power-law ionizing and non-ionizing spectrum, the gas is likely low density, n$_{\rm H}\sim$10 to 100\,\cmthree,\ or low pressure for the isobaric models, P/k$\sim$1000--10000\,K\,\cmthree. We note the isobaric models favor a low metallicity gas of $\approx$1/300 solar. However, we caution that given the simplicity of the modeling, we can only make the broadest of statements that the cloud is likely not solar metallicity and has a diffuse ionization and low intensity \pdr. More observations are certainly needed, but  the data we already have indicates it may be possible to construct photoionization and \pdr\ models for \cgm\ clouds linking the warm ionized gas with the diffuse molecular gas.

\subsubsection{Predictions of modeling: \CII\ and \NII\ emission}

Of course, our simplistic modeling can be used to make testable predictions of the other observable atomic lines in \galaxy.  Two of the most readily observable bright atomic lines in high redshift galaxies are the FIR \CIIline\ and \NIIline\ lines at 158\mum\ and 205\mum,\ respectively. While the excitation mechanisms of the \CII\ line are complex with possible emission from \HII\ regions, PDRs, and shocks \citep[e.g.,][]{Stacey2010, Diaz-Santos2017, Appleton2018}, the \NII\ emission is dominated by \HII\ regions \citep[e.g.,][]{Decarli2014,Zhao2016,Bethermin2016,Zhang2018}. Assuming constant abundance ratios, the [CII]/[NII] ratio can thus provide a constraint on which fraction of the \CII\ flux originates from the \HII\ regions.

The photoionization and \pdr\ models predict that the \CIIline\ and \NIIline\ should be about a factor of several 10--1000 times brighter than the \CIline\ line and that the ratio of \CII/\NII\ should be roughly constant between 1 and 2 (we note that we assumed constant abundance ratios of C to N in our modeling). If our modeling is close to being appropriate for the cloud or clouds we observed, then observing \galaxy\ in both lines would provide a robust test of our model and might constrain the metallicity \citep[e.g.,][]{Nagao2012,Bethermin2016}.
Of course, if the contribution from PDRs to the \CII\ emission is significant, then such observations will be less constraining since some of the emitting gas is not physically related in a simple way. More elaborate Cloudy modeling - including a varying ratio of C and N that depends on the overall metallicity of the gas in the \HII\ and \pdr\ regions, such as what was done by \citet{Nagao2011,Nagao2012} or \citet{Pereira-Santaella2017} - would be required. Comparing the \NII\ emission with the rest-frame UV lines would be appropriate as their emission does originate in the ionization front of the cloud or clouds within this region. Despite these caveats, observing the region in both lines would serve to validate or refute our modeling and potentially constrain both the ionizing intensity and the abundance ratio of C to N, as well as, perhaps, the metallicity. As such, these observations could provide interesting constraints on the \cgm\ of this radio galaxy and a type of constraint that is generally lacking for the \cgm\ of distant galaxies.

\subsubsection{Consistency with the AGN ionization rate}
\label{sec:Consistency_AGN_ionization_rate}

As a consistency check, we estimated the total ionizing energy rate of the \agn\ in the radio galaxy to ensure that it has sufficient energy to actually ionize and excite the cloud. To do this, we used the definition of the ionization parameter, which the Cloudy modeling suggests is approximately 1/1000.  For a cloud that is 75\,kpc from an ionizing source, we find that the required ionizing intensity is 2.1$\times$10$^{55}$ (d$_\mathrm{cloud}$/75\,kpc)$^{2}$ n$_\mathrm{H}$ photons s$^{-1}$. If we assume that all of the ionizing photons have an energy of 13.6\,eV for simplicity, we find that the total ionizing energy of the \agn\ in the radio galaxy has to be at least 4.6$\times$10$^{44}$ (d$_\mathrm{cloud}$/75 kpc)$^{2}$ n$_\mathrm{H}$\,erg\,s$^{-1}$, given our estimated ionization parameter from the modeling. In \citet{Falkendal2019}, we estimated the IR luminosity of the \agn\ to be 4.2$\times$10$^{46}$\,erg\,s$^{-1}$ from SED fitting. The ratio of the IR to UV luminosities is approximately 1 in QSOs \citep[see][]{elvis94, richards06} and thus the IR luminosity estimate puts a rough upper limit on the likely UV luminosity of the \agn. The \agn\ in \galaxy\ only emits sufficient UV photons for  relatively low density gas with n$_{\rm H}\la$100\,\cmthree. If the starburst in \galaxy\ contributes any photons, it can, at most, contribute about a factor of a few more.  However, we would not expect the photons from the recent star formation to escape very far out into the halo of the source. Thus, for the photoionization and \pdr\ model to be applicable, either the gas must have a relatively low density compared to what is normally modeled for PDRs at the surfaces of molecular clouds, or the cloud must lie along a particularly clear line-of-sight since we assumed isotropic QSO emission in our crude calculation.At any rate, it is at least plausible that the \agn\ can excite the cloud.

\subsubsection{Gas masses of the CGM clouds}
\label{sec:molecular_gas_mass_estimate}

Using Eq.~\ref{eqn:MassCI}, we can estimate the masses of molecular gas in the \cgm\ cloud we detected in \CIlower. Assuming the same parameters as we did for the radio galaxy proper (region B), we find
 M$_{\rm H_2}^{\rm north}$=(1.42$\pm$0.95)$\times$10$^{10}$\,\msun\ for the northern detection (region A). We note that we used X$_{\rm [C\textsc{i}]}$=10$^{-5}$, the ratio of \CI\ to \Htwo, to make this estimate. This value is consistent with the value determined in the Milky Way and some distant galaxies \citep[i.e., relatively metal rich,][]{frerking1989,weiss2005,Emonts2018}. However, our simple photoionization modeling is consistent with a significantly lower metallicity in the \cgm\ of \galaxy. Our best estimate of the metallicity is about a factor of ten lower than in the Milky Way, for example, and if this is the case, then the molecular mass would be a factor of ten higher.

The ionized gas mass can be estimated from the \Lya\ emission using the relation from \citet{DeBreuck2003}:

\begin{equation}
{\rm M}_{\rm \HII} = 10^8 \, \big[{\it ff}_{-5}\,L_{44}\,V_{68} \big] ^{1/2}\,{\rm M}_{\odot}
\label{eqn:HIImass}
, \end{equation}

\noindent
where ${\it ff}_{-5}$ is the filling factor in units of $10^{-5}$, $L_{44}$ is the \Lya\ luminosity in units of $10^{44}\,$erg\,s$^{-1}$, and $V_{68}$ is the total volume in units of $10^{68}$\,cm$^3$. We do not know the filling factor.  From the photoionization modeling, we find that densities of about 10-100\,\cmthree\ are consistent with the line ratios. It is simple to show that the volume filling factor assuming case B recombination in the cloud is: 

\begin{equation}
{\it ff}_{\rm V} = V_{\rm em}/V_{\rm geometric}=6.5 \times 10^{-6}\,L_{44}\,n_{100}^{-2}\,V_{68}^{-1} 
\label{eqn:ffV}
, \end{equation}

\noindent
where $V_{\rm em}/V_{\rm geometric}$ is the ratio of the emitting volume and the geometric volume (the geometric volume is defined as the volume over which the \Lya\ emission is located). For regions A and C, we crudely estimated the volume as a cube of 60\,kpc$^3$ (which is equivalent to assuming 2\arcsec\ in projection in three dimensions at the estimated distance of \galaxy); for region B, we used 1\arcsec\ in projection, corresponding to a volume of 7.5\,kpc$^3$. We find that the volume filling factors of the two clouds must be very small. The lower limit and best estimates of the total \Lya\ flux and range of densities suggests ${\it ff}_{\rm V}>10^{-7}$ and up to $\approx$10$^{-4}$ (with the lower value from the lower limits on the \Lya\ fluxes and assuming a density of 100\,cm$^{-3}$). These values are consistent with values found in other studies \citep[$10^{-5}$ to $10^{-6}$;][]{McCarthy1990, Nesvadba2006}. The regions that dominate the emission are composed of cloudlets and rivulets of warm ionized gas that only fill a small fraction of the observed volume. Although the estimates of the masses could have been cast originally as proportional to the density instead of to the volume filling factor, using Eq.~\ref{eqn:HIImass} we estimated the masses of the warm ionized gas in the clouds to be: ${\rm M}_{\rm HII}^{\rm north}$>1.4$\times 10^6$ to 9.4$\times$10$^{7}$\,\msun\ in region A and ${\rm M}_{\rm HII}^{\rm south}$>6.3$\times 10^6$ to 1.3$\times$10$^8$\,\msun\ in region C. The lower limits we used are the lower limits to the \Lya\ fluxes of both regions and the high range of the estimated densities, 100\,\cmthree.  For this particular calculation, we assumed the gas is completely photoionized, that the proton and electron densities are equal, and that much of the intrinsic \Lya\ emission has not been absorbed by dust. We believe we can safely ignore the impact of dust on the \Lya\ fluxes. The low metallicity we estimated for the regions of warm ionized gas and the lack of dust detection indicate that the dust fraction is not high. In addition, a low volume filling factor and velocity differences between the various emission line regions (dispersions of $\approx$130\,\kms) probably mean that each \Lya\ photon is unlikely to encounter many clouds or rivulets along its path out of the region. Thus, scattering is unlikely to be a dominant effect \citep[e.g.,][]{Vernet2017}.

\subsubsection{Importance of multi-phase observations and the nature of the emission region}

Tracing the \cgm\ gas in ionized emission from rest-frame UV lines has revealed large extended halos around high-$z$ galaxies. The \Lya\ emission is seen to be distributed as a more or less uniform sphere around the galaxies \citep{Wisotzki2016,Leclercq2017,leclercq20} or show more complex structures \citep[e.g.,][]{Cantalupo2014, Vernet2017, li19, martin19}. Observing the \cgm\ in emission such as \Lya\ is thus a powerful tool to probe the extent and structure of the \cgm.
On the other hand, the ionized gas only traces warm, very diffuse gas, which is likely not the phase in which most of the gas mass exists. For example, \citet{Emonts2016} and \citet{Emonts2018} show that the cold gas phase in the \cgm\ around the \hzrgs,\ known as the Spiderweb (MRC\,1138-262), contains a large reservoir of molecular gas, \mmol=(1.5$\pm$0.4)$\times$10$^{11}$\,\msun. We find that the northern region of emission (region A) of quiescent \cgm\ gas contains a large amount of molecular gas mass, \mmol$\la$10$^{10}$\,\msun, compared to the ionized gas mass, \mion>$1.3\times 10^{6}$ and up to almost 10$^{8}$\,\msun. The molecular mass is about 2-4 orders of magnitude larger than the ionized gas mass. We note that if the metallicity of region A is as low as we estimated, the molecular gas mass fraction will be even higher (perhaps by up to an order of magnitude). The situation in region C is obviously not so clear.  We do not detect any \CIlower\ emission over this region, which suggests that its molecular gas content is much smaller, while it has a larger ($\sim$3-5 $\times$) \HII\ mass than region A. The limit on the molecular to warm ionized mass is then at least about an order of magnitude smaller in region C than in region A. It is still likely that the molecular gas dominates the mass, but only deeper observations will allow us to determine this quantitatively (the upper limit for the difference is still
of the order of three orders of magnitude!).

Empirically, it appears that the \cgm\ is best observed in molecular lines, as it may probe the phase that dominates the mass budget. Most of the work on high-$z$ galaxies have so far studied either the ionized gas or molecular gas, but rarely both. Complementing ionized emission studies with molecular line tracers provides a novel way to probe the complex multi-phase nature of the \cgm. It may also probe a larger fraction of the gas mass of the \cgm\ and the morphology of the total mass distribution of the system, especially the cores in the mass distribution \citep[see][]{emonts19}.

\begin{table}
\centering
\begin{tabular}{+l ^c ^c ^c}
\toprule
\toprule
Mass    & North      & Core             & South \\
        & Region A   & Region B     & Region C \\
                & \msun\     & \msun\       & \msun \\
\midrule
\mstar\ & \nodata    & 1.3$\times 10^{11}$      & \nodata\      \\
\mion$ (\rm min)^*$  & $>$1.4$\times$10$^6$ & $>$1.0$\times$10$^7$ & $>$6.3$\times$10$^6$ \\
\mion$ (\rm max)^{\dagger}$  & 9.4$\times$10$^7$ & 1.8$\times$10$^8$ & 1.3$\times$10$^8$ \\
\mmol\  & $\la$1.4$\times$10$^{10}$ & 3.1$\times$10$^{10}$ & $<$1.7$\times$10$^{9}$\\
\bottomrule
\end{tabular}
\caption{Stellar mass and ionized- and molecular-gas mass for the host galaxy (region B) and the diffuse extended north component (region A) detected in \CIV\ and \CIlower. $^*$Lower limits estimated by not considering in the line fits any part of the \Lya\ profile that was impacted by strong night sky line emission or possible associated absorption and a density of 100 cm$^{-3}$. $^{\dagger}$The mass estimate of the \mion\ comes from the fit of the line profile corrected for associated HI absorption and assuming a density of 10 cm$^{-3}$. Assuming a lower density results in higher \mion\ masses. See text for details on how the densities are constrained.}
\label{Tab:variousprops}
\end{table}

Our modeling suggests, since circumgalactic gas is estimated to have much lower average densities, that this cloud of gas is either transient (in that it has a pressure higher than that of the surrounding medium) or it is pressure confined and longer lived.  In Sect.~\ref{subsec:photomod}, for our isobaric models, we estimated pressures of P/k$\sim$1000--10000\,K\,\cmthree\ in the cloud. For the cloud to be pressure confined, assuming that there is a hot halo with a temperature approximately that of the virial temperature (a few 10$^{6-7}$\,K for a halo with a mass of 10$^{13-14}$\,\msun), it would require densities of $\sim$10$^{-3}$\,\cmthree. At such large radii in the halo, it is not clear if there is sufficient pressure to confine clouds like the ones we discovered in this study.  While we do not know the mass of the halo, the high stellar mass (Table~\ref{Tab:variousprops}) likely indicates that it is massive \citep[$\sim10^{13}$\,\msun;][]{Legrand2019}. Numerical simulations suggest that there is sufficient thermal pressure in massive halos at z=3 to confine clouds of the properties we have estimated \citep{Rosdahl2012}. If this is the case, then observing the emission from the warm ionized gas that is offset from the molecular gas suggests that the cloud is being photo-evaporated and eroded by the momenta of the photons striking and heating the dense gas.

The dynamics and spatial offset between the rest-frame UV and submm line emission may suggest that the warm ionized gas and the diffuse molecular gas are not related in contrast to the photoionization and \pdr\ modeling. The explanation, however, may lie in the differences in the nature of the ionization fronts. We do not have enough detailed observations to know for sure, but perhaps the best analogy for this situation is a cloud of gas that is being photo-evaporated, such as the ``Pillars of Creation'' in M16 \citep{McLeod2015} or in the pillars of dense gas near NGC 3603 \citep{Westmoquette2013} in our Galaxy. In these photo-evaporation flows, the highly ionized gas is accelerated at the edge of the denser molecular pillar by the intense radiation field responsible for ionizing the gas at the photo-evaporation front, from which it is spatially offset.  Over time, the cloud will be completely dispersed by the intense radiation field.  While perhaps not a perfect analogy, it does provide the sense that the warm ionized gas gets accelerated as well as ionized, and that it reaches both higher thermal and ram pressures compared to the quiescent dense diffuse molecular gas. In addition, since it is accelerating and flowing away, the warm ionized and molecular gas phases are no longer perfectly co-spatial. Such a picture may explain the differences in the kinematics and the more extended, offset morphology of the UV emission lines while still being consistent with a very simplistic photoionization and photon-dominated model of a cloud of gas.

\section{Conclusions}

We observed the \hzrg\ \galaxy\ in rest-frame UV emission lines with MUSE and in \CIlower\ with ALMA. The combination allowed us to probe the ionized gas and the molecular gas of the host galaxy, as well as of the surrounding \cgm.

\begin{itemize}

\item We detect narrow, dynamically quiescent \CIlower\ at the core of the host galaxy. In order to explain the very narrow line width, we need to be viewing the galaxy face-on, rotating in the plane of the sky. This would mean that the host galaxy is not aligned with the nuclear launching region, as we are viewing this type~II \agn\ edge-on, and the radio jets are propagating in the plane of the sky.

\item We weakly detect \CIlower\ emission $\sim$75\,kpc away from the host galaxy. This emission is not perfectly co-spatial but does lie approximately within the region of weak extended \CIV\ emission. The two carbon lines do not have the same \fwhm\ and velocity offset relative to the systemic redshift, but they are both very narrow and must originate from dynamically cold gas. 

\item We performed photoionization and photon-dominated region modeling using Cloudy to investigate the possible nature of the extended quiescent gas detected in \CIV\ and \CI. We are able to explain the observed \CI/\CIV\ and \CIV/\HeII\ flux ratios with simple assumptions, and they are consistent with \pdr-dominated regions in the \cgm. We stress that this is meant as a proof of concept and not intended to be an exhaustive analysis.

\item The photoionization and \pdr\ modeling suggests that the observed multi-phase region has a low metallicity, low ionization, and low density. We show that the luminosity of the \agn\ is sufficient to ionize the gas out to 75\,kpc if the gas is diffuse enough (i.e.,  n$_H\la$100\,\cmthree). This is consistent with the idea that these regions of the \cgm\ are likely being photo-evaporated and eroded by the momenta of the photons striking and heating the dense gas observed in \CI.

\end{itemize}

Our data suggest that the \cgm\ in \galaxy\ contains a multi-phase region of ionized and molecular gas. This gas cloud can be explained as being ionized by the \agn. We are limited to low signal-to-noise and the source is at an unfortunate redshift, which makes the \Lya\ and \HeII\ unusable for kinematic studies. We therefore rely only on the weak \CIV\ emission. The \CI\ has been shown to probe a larger amount of gas mass than what ionization lines can trace. This work shows the possibilities and power of observing galaxies using both ionized and molecular gas. We would need to observe \galaxy\ in more lines to get a deeper understanding of the physical condition of the \cgm. The \hzrgs\ provide unique opportunities to study the \cgm, since they are luminous enough to excite the gas out to large distances. The radio jets and ionization cones act as flashlights, illuminating the surrounding diffuse \cgm\ and making it shine.

\begin{acknowledgements}
We thank Eric Emsellem and Guy Perrin for their help in securing financial support for this research and the work of TF. TF and MDL would like to thank Fiorella Polles for helpful discussions on the use and interpretation of Cloudy and MDL thanks Pierre Guillard for advice and interesting discussions about \pdr\ physics and modeling. TF would like to thank Susanne Aalto on her valuable input on an earlier version of this paper. This study makes use of ADS/JAO.ALMA\#2015.1.00530.S.
ALMA is a partnership of ESO (representing its member states), NSF (USA) and NINS (Japan), together with NRC
(Canada), NSC and ASIAA (Taiwan), and KASI (Republic of Korea), in
cooperation with the Republic of Chile. This work fulfills part of the Ph.D. requirements for Theresa Falkendal at Sorbonne Universit\'{e}. We thank the anonymous referee for their comments which lead to an overall better manuscript.
\end{acknowledgements}

\bibliographystyle{aa}
\bibliography{References}

\clearpage
\newpage

\begin{appendix}

\section{Redshift of foreground sources}
\label{sec:redshift_of_foreground}
In the field-of-view of MUSE around \galaxy,\ there are four foreground galaxies: two of which have previously been presented in literature with counterparts in other wavelengths and two are, to our knowledge, new or previously unpublished. Table~\ref{tab:coordinates} summarizes the detected lines, redshifts, and coordinates of the galaxies shown in Fig.~\ref{fig:support_data}.

\begin{figure}[ht]
\includegraphics[width=\linewidth]{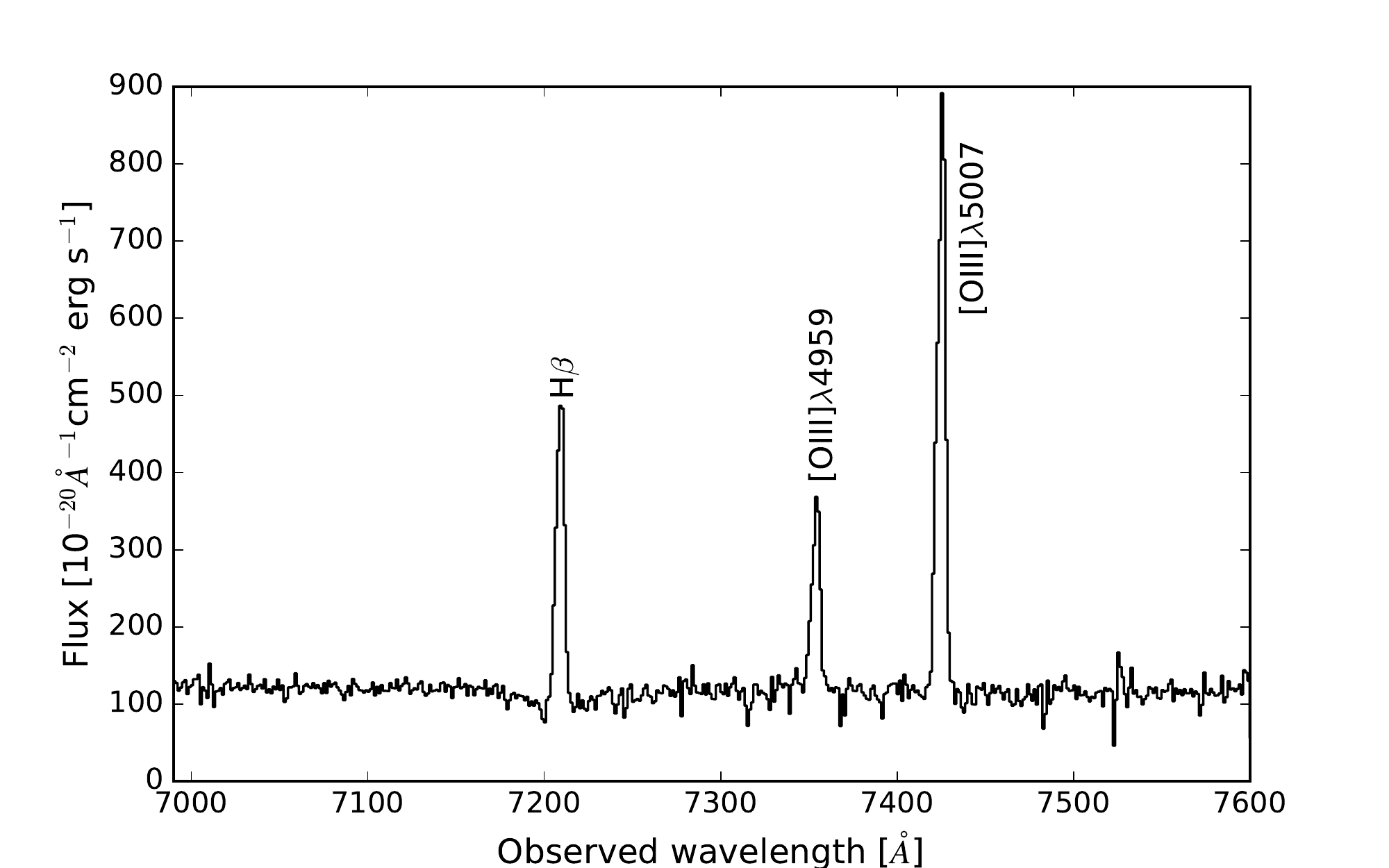}
\caption{The spectrum of the foreground of Galaxy 1 at z=0.483. The spectrum was extracted over a circular aperture with a radius of 3 pixels. We detected both the \OIII$\lambda\lambda$4959,5007 doublet and H$\beta$ line.}
\label{fig:Galaxy1_spectra}
\end{figure}

\begin{figure}[ht]
\includegraphics[width=\linewidth]{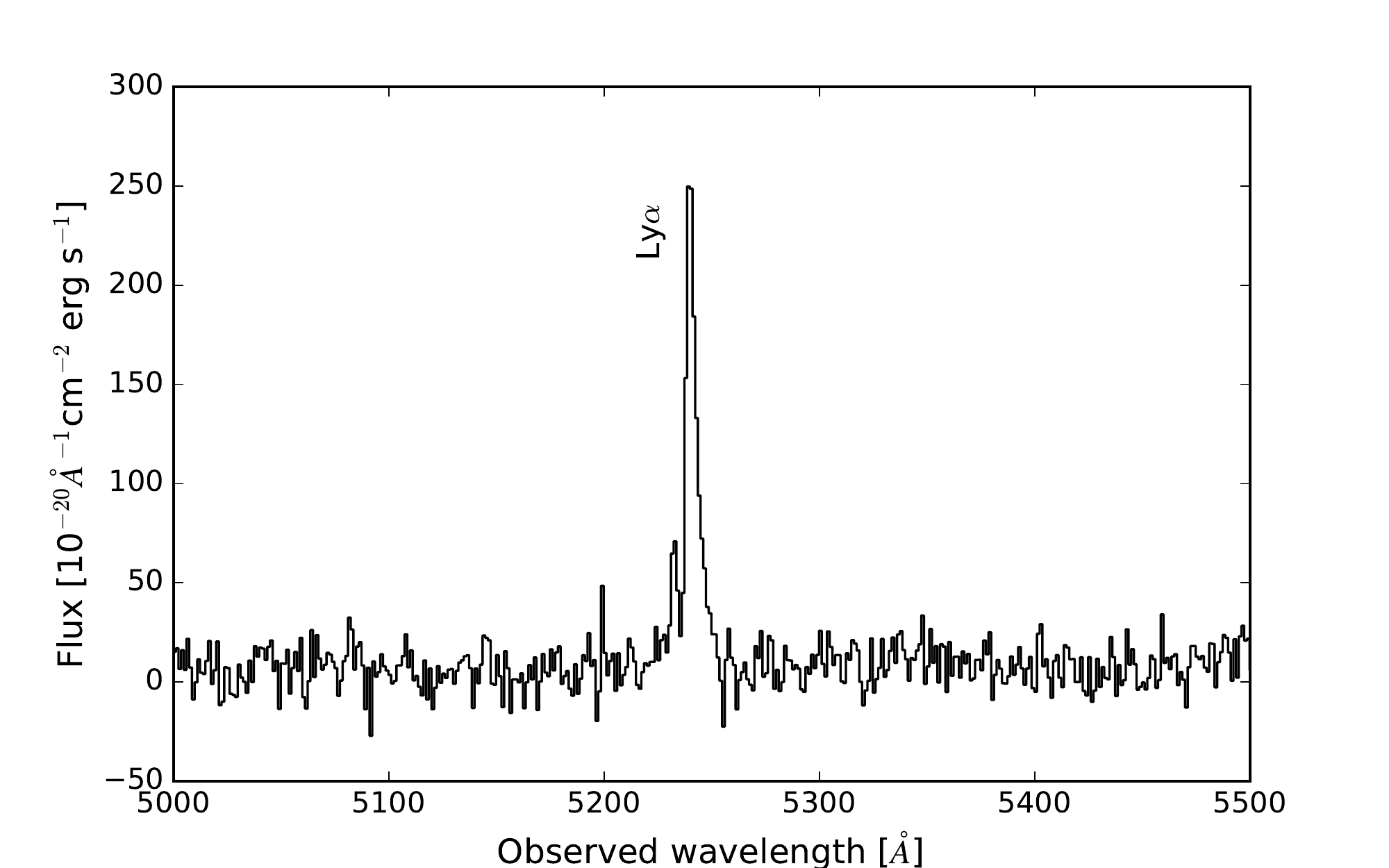}
\caption{The spectrum of the foreground of Galaxy 2 at z=3.31. The spectrum was extracted over a circular aperture with a radius of 3 pixels. We assumed the only line detected is due to \Lya\ emission, but it is not possible to estimate a robust redshift without a detection in any other lines.}
\label{fig:Galaxy2_spectra}
\end{figure}

\begin{figure}[ht]
\includegraphics[width=\linewidth]{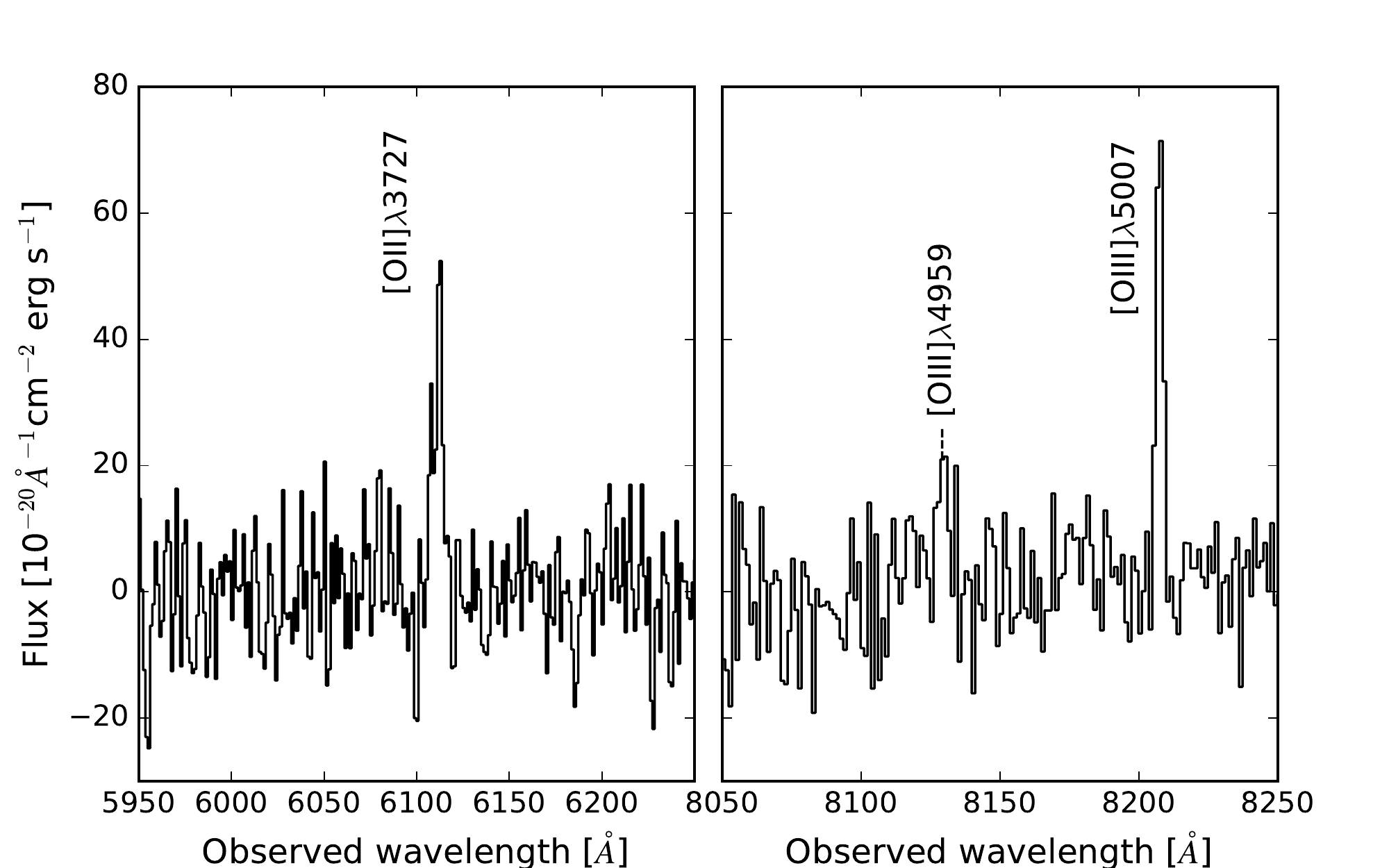}
\caption{The spectrum of the foreground of Galaxy 3 at a redshift of z=0.639. The spectrum was extracted over a circular aperture with a radius of 3 pixels. We detected the \OII$\lambda$3726,3729 and \OIII$\lambda\lambda$4959,5007 doublets.}
\label{fig:Galaxy3_spectra}
\end{figure}

Galaxy 1 has been detected in X-ray, K-band, and IRAC 1 and now with MUSE in the emission lines H$\beta,$ and \OIII$\lambda\lambda$4959,5007 (Fig.~\ref{fig:Galaxy1_spectra}). From these three lines, we measure a redshift of z=0.483. Galaxy 2 is only detected in one line, which we assume is \Lya\ (Fig.~\ref{fig:Galaxy2_spectra}). If the line is indeed \Lya,\ then the redshift of the source is z=3.31. Although it is only detected in one emission line, it is likely that this is a high-$z$ source since it is not seen in any of the previous observations. Galaxy 3 is not observed in any of the photometric bands, but it is detected in the \OII$\lambda$3726,3729 and \OIII$\lambda\lambda$4959,5007 lines (Fig.~\ref{fig:Galaxy3_spectra}) and has a redshift of z=0.639. The fourth foreground source, Galaxy 4, is detected in K-band and IRAC 1; the MUSE spectra show only one strong line. We interpret this as \OII$\lambda$3727 since the line coincides with a continuum break, likely to be the 4000\,\AA\ break with D$_{4000}=1.6$, where D is the flux density ratio between the wavelength ranges 4100--4400\,\AA\ and 3600--3900\,\AA.

\begin{figure}[ht]
\includegraphics[width=\linewidth]{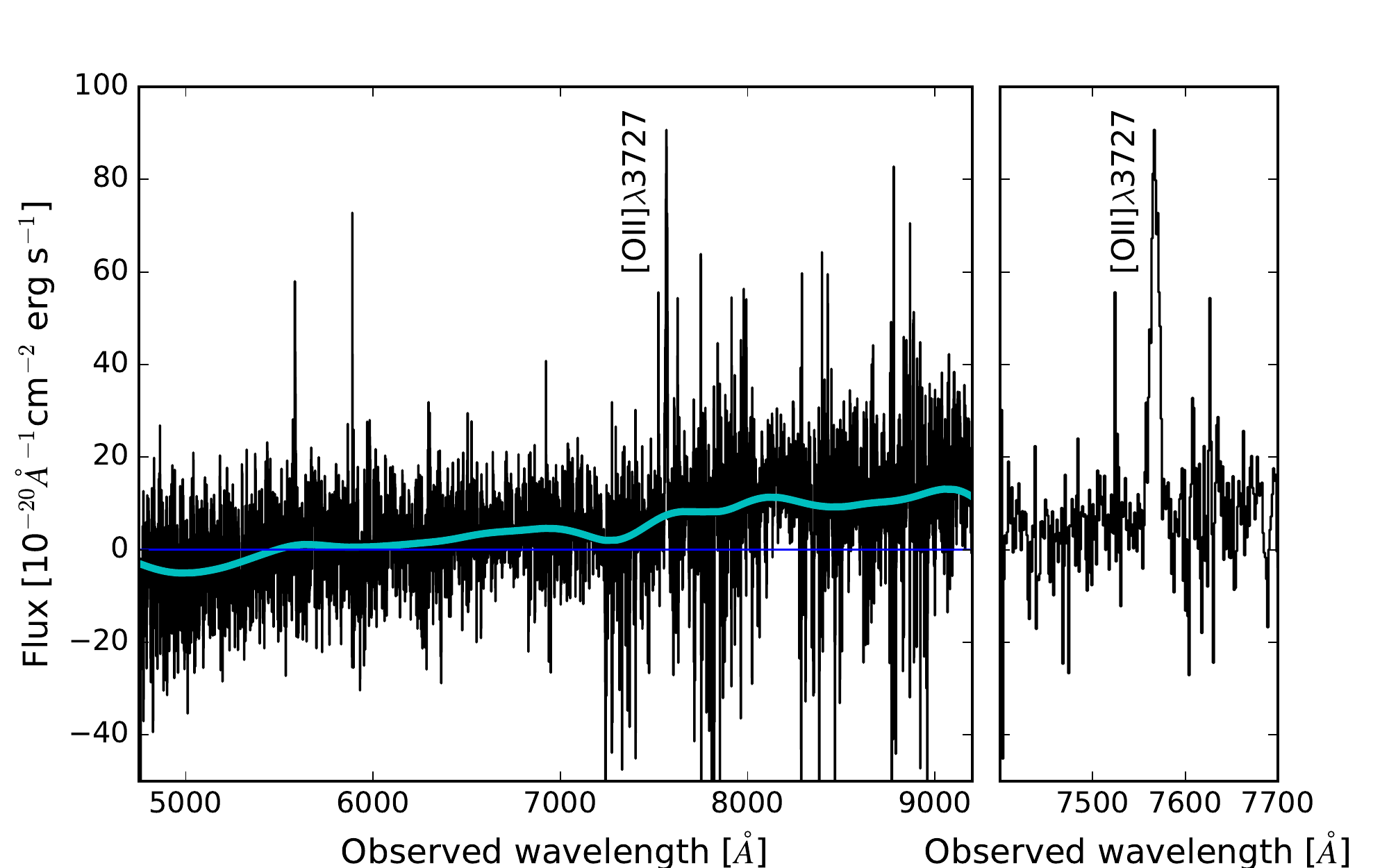}
\caption{The spectrum of the foreground of Galaxy 4 at a redshift of z=1.03. We extracted the spectrum over a circular aperture with a radius of 3 pixels. \textit{Left panel:} The full spectrum of the extracted region, which is contaminated with \Lya\ emission from \galaxy\ at $\lambda_{\rm obs}\sim$5580\,\AA. The cyan line shows the spectrum after it was Hanning-smoothed and the blue line indicates the zero level. \textit{Right panel:} A zoom-in of the line at $\lambda_{\rm obs}$=7567.95\,\AA\ in the spectrum of Galaxy 4. The redshift is determined as if the detection were the \OII\ line. Such an identification is consistent with a continuum break observed in the spectrum. We believe this is likely the 4000\,\AA\ break, which is consistent with our line identification.}
\label{fig:Galaxy4_spectra}
\end{figure}

\begin{table*}
\centering
\begin{tabular}{lcccc}
\toprule
Source          & $\alpha$(J2000)   & $\delta$(J2000)& redshift & line of which $z$ is determined \\
\midrule
\galaxy$^{*}$                   & 21:44:07.52           & 19:29:14.2    & 3.5895 & \CI    \\
North radio hot spot            & 21:44:07.49           & 19:29:19.1    &\nodata        &\nodata\\
South radio hot spot            & 21:44:07.53           & 19:29:10.8            &\nodata        &\nodata\\
Galaxy 1                                & 21:44:07.67           & 19:29:07.7         &0.483  & H$\beta$, \OIII$\lambda$4959, \OIII$\lambda$5007 \\
Galaxy 2                                & 21:44:07.47           & 19:29:05.7    & 3.31$^u$        &\Lya $\lambda$1215.7\\
Galaxy 3                                & 21:44:07.56           & 19:29:11.1    & 0.639 &\OII$\lambda$3727, \OIII$\lambda$4959, \OIII$\lambda$5007\\
Galaxy 4                                & 21:44:07.36           & 19:29:19.3    & 1.03  &\OII$\lambda$3727\\
\bottomrule                          
\end{tabular}
\caption{Coordinates of sources in the field and redshift of the foreground objects. ($*$) Center of the host galaxy determined from the peak of the thermal dust emission of the ALMA band 3 continuum image. ($u$) Redshift only estimated from one line and is thus uncertain.}
\label{tab:coordinates}
\end{table*}

\end{appendix}

\end{document}